\input harvmac
\input epsf
%
\newbox\hdbox%
\newcount\hdrows%
\newcount\multispancount%
\newcount\ncase%
\newcount\ncols
\newcount\nrows%
\newcount\nspan%
\newcount\ntemp%
\newdimen\hdsize%
\newdimen\newhdsize%
\newdimen\parasize%
\newdimen\spreadwidth%
\newdimen\thicksize%
\newdimen\thinsize%
\newdimen\tablewidth%
\newif\ifcentertables%
\newif\ifendsize%
\newif\iffirstrow%
\newif\iftableinfo%
\newtoks\dbt%
\newtoks\hdtks%
\newtoks\savetks%
\newtoks\tableLETtokens%
\newtoks\tabletokens%
\newtoks\widthspec%
%
%
%
%
\tableinfotrue%
\catcode`\@=11
%
%
\def\tstrut{\vrule height3.1ex depth1.2ex width0pt}%
\def\and{\char`\&}
\def\tablerule{\noalign{\hrule height\thinsize depth0pt}}%
\thicksize=1.5pt
\thinsize=0.6pt
\def\thickrule{\noalign{\hrule height\thicksize depth0pt}}%
\def\ctr#1{\hfil\ #1\hfil}%
%
%
%
%
\tablewidth=-\maxdimen%
\spreadwidth=-\maxdimen%
\def\tabskipglue{0pt plus 1fil minus 1fil}%
%
%
\centertablestrue%
%
%
%
%
\parasize=4in%
\gdef\ARGS{########}
\gdef\headerARGS{####}
\def\@mpersand{&}
{\catcode`\|=13
\gdef\letbarzero{\let|0}
\gdef\letbartab{\def|{&&}}%
\gdef\letvbbar{\let\vb|}%
}
{\catcode`\&=4
\def\ampskip{&\omit\hfil&}
\catcode`\&=13
\let&0
\xdef\letampskip{\def&{\ampskip}}%
\gdef\letnovbamp{\let\novb&\let\tab&}
}
\def\begintable{
   \begingroup%
   \catcode`\|=13\letbartab\letvbbar%
   \catcode`\&=13\letampskip\letnovbamp%
   \def\multispan##1{
      \omit \mscount##1%
      \multiply\mscount\tw@\advance\mscount\m@ne%
      \loop\ifnum\mscount>\@ne \sp@n\repeat%
   }
   \def\|{%
      &\omit\widevline&%
   }%
   \ruledtable
}
\long\def\ruledtable#1\endtable{%
%
%
%
   \offinterlineskip
   \tabskip 0pt
   \def\widevline{\vrule width\thicksize}
   \def\endrow{\@mpersand\omit\hfil\crnorm\@mpersand}%
   \def\crthick{\@mpersand\crnorm\thickrule\@mpersand}%
   \def\crthickneg##1{\@mpersand\crnorm\thickrule
          \noalign{{\skip0=##1\vskip-\skip0}}\@mpersand}%
   \def\crnorule{\@mpersand\crnorm\@mpersand}%
   \def\crnoruleneg##1{\@mpersand\crnorm
          \noalign{{\skip0=##1\vskip-\skip0}}\@mpersand}%
   \let\nr=\crnorule
   \def\endtable{\@mpersand\crnorm\thickrule}%
   \let\crnorm=\cr
%
%
   \edef\cr{\@mpersand\crnorm\tablerule\@mpersand}%
   \def\crneg##1{\@mpersand\crnorm\tablerule
          \noalign{{\skip0=##1\vskip-\skip0}}\@mpersand}%
   \let\ctneg=\crthickneg
   \let\nrneg=\crnoruleneg
   \the\tableLETtokens
%
%
   \tabletokens={&#1}
%
%
   \countROWS\tabletokens\into\nrows%
   \countCOLS\tabletokens\into\ncols%
%
%
   \advance\ncols by -1%
   \divide\ncols by 2%
   \advance\nrows by 1%
%
%
   \iftableinfo %
      \immediate\write16{[Nrows=\the\nrows, Ncols=\the\ncols]}%
   \fi%
%
%
   \ifcentertables
      \ifhmode \par\fi
      \line{
      \hss
   \else %
      \hbox{%
   \fi
      \vbox{%
         \makePREAMBLE{\the\ncols}
         \edef\next{\preamble}
         \let\preamble=\next
         \makeTABLE{\preamble}{\tabletokens}
      }
      \ifcentertables \hss}\else }\fi
   \endgroup
   \tablewidth=-\maxdimen
   \spreadwidth=-\maxdimen
}
\def\makeTABLE#1#2{
   {
   \let\ifmath0
   \let\header0
   \let\multispan0
%
%
   \ncase=0%
   \ifdim\tablewidth>-\maxdimen \ncase=1\fi%
   \ifdim\spreadwidth>-\maxdimen \ncase=2\fi%
   \relax
%
   \ifcase\ncase %
      \widthspec={}%
   \or %
      \widthspec=\expandafter{\expandafter t\expandafter o%
                 \the\tablewidth}%
   \else %
      \widthspec=\expandafter{\expandafter s\expandafter p\expandafter r%
                 \expandafter e\expandafter a\expandafter d%
                 \the\spreadwidth}%
   \fi %
   \xdef\next{
      \halign\the\widthspec{%
      #1
      \noalign{\hrule height\thicksize depth0pt}
      \the#2\endtable
%
      }
   }
   }
   \next
}
\def\makePREAMBLE#1{
   \ncols=#1
   \begingroup
   \let\ARGS=0
   \edef\xtp{\widevline\ARGS\tabskip\tabskipglue%
   &\ctr{\ARGS}\tstrut}
   \advance\ncols by -1
   \loop
      \ifnum\ncols>0 %
      \advance\ncols by -1%
      \edef\xtp{\xtp&\vrule width\thinsize\ARGS&\ctr{\ARGS}}%
   \repeat
   \xdef\preamble{\xtp&\widevline\ARGS\tabskip0pt%
   \crnorm}
   \endgroup
}
\def\countROWS#1\into#2{
   \let\countREGISTER=#2%
   \countREGISTER=0%
   \expandafter\ROWcount\the#1\endcount%
}%
\def\ROWcount{%
   \afterassignment\subROWcount\let\next= %
}%
\def\subROWcount{%
   \ifx\next\endcount %
      \let\next=\relax%
   \else%
      \ncase=0%
      \ifx\next\cr %
         \global\advance\countREGISTER by 1%
         \ncase=0%
      \fi%
      \ifx\next\endrow %
         \global\advance\countREGISTER by 1%
         \ncase=0%
      \fi%
      \ifx\next\crthick %
         \global\advance\countREGISTER by 1%
         \ncase=0%
      \fi%
      \ifx\next\crnorule %
         \global\advance\countREGISTER by 1%
         \ncase=0%
      \fi%
      \ifx\next\crthickneg %
         \global\advance\countREGISTER by 1%
         \ncase=0%
      \fi%
      \ifx\next\crnoruleneg %
         \global\advance\countREGISTER by 1%
         \ncase=0%
      \fi%
      \ifx\next\crneg %
         \global\advance\countREGISTER by 1%
         \ncase=0%
      \fi%
      \ifx\next\header %
         \ncase=1%
      \fi%
      \relax%
      \ifcase\ncase %
         \let\next\ROWcount%
      \or %
         \let\next\argROWskip%
      \else %
      \fi%
   \fi%
   \next%
}
\def\counthdROWS#1\into#2{%
\dvr{10}%
   \let\countREGISTER=#2%
   \countREGISTER=0%
\dvr{11}%
\dvr{13}%
   \expandafter\hdROWcount\the#1\endcount%
\dvr{12}%
}%
\def\hdROWcount{%
   \afterassignment\subhdROWcount\let\next= %
}%
\def\subhdROWcount{%
   \ifx\next\endcount %
      \let\next=\relax%
   \else%
      \ncase=0%
      \ifx\next\cr %
         \global\advance\countREGISTER by 1%
         \ncase=0%
      \fi%
      \ifx\next\endrow %
         \global\advance\countREGISTER by 1%
         \ncase=0%
      \fi%
      \ifx\next\crthick %
         \global\advance\countREGISTER by 1%
         \ncase=0%
      \fi%
      \ifx\next\crnorule %
         \global\advance\countREGISTER by 1%
         \ncase=0%
      \fi%
      \ifx\next\header %
         \ncase=1%
      \fi%
\relax%
      \ifcase\ncase %
         \let\next\hdROWcount%
      \or%
         \let\next\arghdROWskip%
      \else %
      \fi%
   \fi%
   \next%
}%
{\catcode`\|=13\letbartab
\gdef\countCOLS#1\into#2{%
   \let\countREGISTER=#2%
   \global\countREGISTER=0%
   \global\multispancount=0%
   \global\firstrowtrue
   \expandafter\COLcount\the#1\endcount%
   \global\advance\countREGISTER by 3%
   \global\advance\countREGISTER by -\multispancount
}%
\gdef\COLcount{%
   \afterassignment\subCOLcount\let\next= %
}%
{\catcode`\&=13%
\gdef\subCOLcount{%
   \ifx\next\endcount %
      \let\next=\relax%
   \else%
      \ncase=0%
      \iffirstrow
         \ifx\next& %
            \global\advance\countREGISTER by 2%
            \ncase=0%
         \fi%
         \ifx\next\span %
            \global\advance\countREGISTER by 1%
            \ncase=0%
         \fi%
         \ifx\next| %
            \global\advance\countREGISTER by 2%
            \ncase=0%
         \fi
         \ifx\next\|
            \global\advance\countREGISTER by 2%
            \ncase=0%
         \fi
         \ifx\next\multispan
            \ncase=1%
            \global\advance\multispancount by 1%
         \fi
         \ifx\next\header
            \ncase=2%
         \fi
         \ifx\next\cr       \global\firstrowfalse \fi
         \ifx\next\endrow   \global\firstrowfalse \fi
         \ifx\next\crthick  \global\firstrowfalse \fi
         \ifx\next\crnorule \global\firstrowfalse \fi
         \ifx\next\crnoruleneg \global\firstrowfalse \fi
         \ifx\next\crthickneg  \global\firstrowfalse \fi
         \ifx\next\crneg       \global\firstrowfalse \fi
      \fi
\relax
      \ifcase\ncase %
         \let\next\COLcount%
      \or %
         \let\next\spancount%
      \or %
         \let\next\argCOLskip%
      \else %
      \fi %
   \fi%
   \next%
}%
\gdef\argROWskip#1{%
   \let\next\ROWcount \next%
}
\gdef\arghdROWskip#1{%
   \let\next\ROWcount \next%
}
\gdef\argCOLskip#1{%
   \let\next\COLcount \next%
}
}
}
\def\spancount#1{
   \nspan=#1\multiply\nspan by 2\advance\nspan by -1%
   \global\advance \countREGISTER by \nspan
   \let\next\COLcount \next}%
\def\dvr#1{\relax}%
\def\header#1{%
\dvr{1}{\let\cr=\@mpersand%
\hdtks={#1}%
\counthdROWS\hdtks\into\hdrows%
\advance\hdrows by 1%
\ifnum\hdrows=0 \hdrows=1 \fi%
\dvr{5}\makehdPREAMBLE{\the\hdrows}%
\dvr{6}\getHDdimen{#1}%
{\parindent=0pt\hsize=\hdsize{\let\ifmath0%
\xdef\next{\valign{\headerpreamble #1\crnorm}}}\dvr{7}\next\dvr{8}%
}%
}\dvr{2}}
\def\makehdPREAMBLE#1{
\dvr{3}%
\hdrows=#1
{
\let\headerARGS=0%
\let\cr=\crnorm%
\edef\xtp{\vfil\hfil\hbox{\headerARGS}\hfil\vfil}%
\advance\hdrows by -1
\loop
\ifnum\hdrows>0%
\advance\hdrows by -1%
\edef\xtp{\xtp&\vfil\hfil\hbox{\headerARGS}\hfil\vfil}%
\repeat%
\xdef\headerpreamble{\xtp\crcr}%
}
\dvr{4}}
\def\getHDdimen#1{%
\hdsize=0pt%
\getsize#1\cr\end\cr%
}
\def\getsize#1\cr{%
\endsizefalse\savetks={#1}%
\expandafter\lookend\the\savetks\cr%
\relax \ifendsize \let\next\relax \else%
\setbox\hdbox=\hbox{#1}\newhdsize=1.0\wd\hdbox%
\ifdim\newhdsize>\hdsize \hdsize=\newhdsize \fi%
\let\next\getsize \fi%
\next%
}%
\def\lookend{\afterassignment\sublookend\let\looknext= }%
\def\sublookend{\relax%
\ifx\looknext\cr %
\let\looknext\relax \else %
   \relax
   \ifx\looknext\end \global\endsizetrue \fi%
   \let\looknext=\lookend%
    \fi \looknext%
}%
%
%
\def\tablelet#1{%
   \tableLETtokens=\expandafter{\the\tableLETtokens #1}%
}%
\catcode`\@=12

\def\NSVZ{{\rm NSVZ}} 
\def\DRED{{\rm DRED}}
\def\DREDp{{\rm DRED}'}

\def\half{{\textstyle{{1}\over{2}}}}
\def\frak#1#2{{\textstyle{{#1}\over{#2}}}}
\def\frakk#1#2{{{#1}\over{#2}}}
\def\fivebar{{\overline{5}}}
\def\tenbar{{\overline{10}}}
\def\ga{\gamma}

\def\Ocal{{\cal O}}
\def\tautil{\tilde\tau}

\def\nutil{\tilde\nu}

\def\btil{\tilde b}

\def\dtil{\tilde d}
\def\etil{\tilde e}
\def\gtil{\tilde g}

\def\ttil{\tilde t}
\def\util{\tilde u}

\def\Btil{\tilde B}

\def\Etil{\tilde E}
\def\Ttil{\tilde T}
\def\Ytil{\tilde Y}
\def\sy{supersymmetry}
\def\sic{supersymmetric}

\def\smgroup{$SU_3\otimes\ SU_2\otimes\ U_1$}

\def\npb{{Nucl.\ Phys.\ }{\bf B}}
\def\plb{{Phys.\ Lett.\ }{ \bf B}}

\def\prd{{Phys.\ Rev.\ }{\bf D}}

\def\lf{16\pi^2}
\def\llf{(16\pi^2)^2}

\def\TeV{{\rm TeV}}
\def\GeV{{\rm GeV}}

\thicksize=0.7pt
\thinsize=0.5pt
\def\ctr#1{\hfil $\,\,\,#1\,\,\,$ \hfil}
\def\tstrut{\vrule height 2.7ex depth 1.0ex width 0pt}

\def \inparg{\leftskip = 40pt\rightskip = 40pt}
\def \outparg{\leftskip = 0 pt\rightskip = 0pt}
\def\msbar{{\overline{\rm MS}}}

{\nopagenumbers
\line{\hfil LTH 628}
\line{\hfil CERN-PH-TH/2004-153}
\line{\hfil hep-ph/0408128}
\vskip .5in
\centerline{\titlefont Snowmass Benchmark Points 
and Three-Loop Running} 
\vskip 1in
\centerline{\bf I.~Jack, D.R.T.~Jones\foot{address from Sept 1st 2003-
31 Aug 2004: TH Division, CERN, 1211 Geneva 23, Switzerland} and A.F.~Kord}
\bigskip
\centerline{\it Department of Mathematical Sciences, 
University of Liverpool, Liverpool L69 3BX, U.K.}
\vskip .3in

We present the full three-loop $\beta$-functions for the MSSM
generalised to include additional matter multiplets in 5, 10 
representations of SU(5). We analyse  
the effect of three-loop running on 
the sparticle spectrum for the MSSM Snowmass Benchmark Points. 
We also consider the  effect on these spectra of additional 
matter multiplets (the semi-perturbative unification scenario).

\Date{August 2004}}  

\newsec{Introduction}

The LHC will soon resolve the question as to 
whether low energy \sy\ is the solution to the hierarchy problem; and 
if it is, moreover, 
the LHC and a future $e^+ e^-$ linear collider (LC) will 
lead to very precise measurements of the sparticle spectrum and 
couplings. The success of gauge unification in the MSSM 
suggests a Desert, the existence of which would mean that  extrapolation
of the MSSM couplings and masses to high scales will lead to 
immediate information about the underlying theory; for example 
regarding the commonly assumed universality of soft scalar masses, 
gaugino masses and cubic scalar interactions. 

One component of this analysis is the running of masses and couplings 
between the weak and gauge unification scales, which is governed by the
renormalisation group $\beta$-functions. In this paper we compare 
the results for this  process using one, two and three-loop 
$\beta$-functions. In each case we generally use the same one-loop
corrections  for the relationship between running and pole masses for 
the various particles, with some use of two-loop results such  as for
the top quark mass. We anticipate that  by the time sparticles are
discovered  complete two-loop threshold corrections will be available; 
the effect of these we would expect to be of the same order  of
magnitude as the effect of using the three-loop (as opposed to two-loop)
 $\beta$-functions, which, as we shall see, is  surprisingly large for
squarks. 

The plan of this paper is as follows. In section~2 we review the 
exact results that relate the $\beta$-functions for the soft masses and 
interactions\ref\jjpa{I.~Jack and   D.R.T.~Jones 
\plb415 (1997) 383 }%
\nref\jjpb{I.~Jack, D.R.T.~Jones and A.~Pickering,  
\plb432 (1997) 114}
--\ref\akk{L.V.~Avdeev, D.I.~Kazakov and I.N.~Kondrashuk, 
\npb510 (1998) 289}\ 
to the $\beta$-functions of the dimensionless gauge and Yukawa couplings
\ref\JackQQ{
I.~Jack, D.R.T.~Jones and C.G.~North,
\npb473 (1996) 308
}%
\nref\JackCN{
I.~Jack, D.R.T.~Jones and C.G.~North,
\npb486 (1997) 479 
}--
\ref\fjja{P.M.~Ferreira, I.~Jack and D.R.T.~Jones,
\plb387 (1996) 80 
}, which we then give through three loops for the MSSM generalised to 
incorporate $n_5$ and $n_{10}$ sets of $SU_5$  $5(\fivebar)$ and $10(\tenbar)$
representations respectively. (A motive for grouping additional 
matter in this way is that complete $SU_5$ representations do not 
(at one loop) change the prediction of $\sin^2\theta_W$ (or alternatively 
of $g_3^2 (M_Z)$)  that follows 
from imposing $g_{1,2,3}$ gauge unification. Also unchanged 
at one loop is the gauge unification scale, $M_X$; 
but at higher loops this scale increases and can 
approach the string scale.) We also give a simplified  example of 
a three-loop soft $\beta$-function; general results 
for all the $\beta$-functions are available at 
Ref.~\ref\website{http//www.liv.ac.uk/$\sim$dij/betas}.

In section 3 we present and discuss our results for the 
sparticle spectrum for a set of 
Snowmass Benchmark Points\ref\AllanachNJ{
B.C.~Allanach {\it et al.},
Eur.\ Phys.\ J.\ {\bf C} 25 (2002) 113 
},
all corresponding to the standard universal boundary 
conditions at unification, except for one case with non-universal 
gaugino masses.  We compare our results with 
the   useful website  
Ref.~\ref\sabine{http://kraml.home.cern.ch/kraml/comparison} 
(see also Refs.~\ref\AllanachJW{
B.C.~Allanach, S.~Kraml and W.~Porod,
JHEP  016 (2003) 0303
}, \ref\GKG{N.~Ghodbane and H.U.~Martyn, hep-ph/0201233.}
).

In section~4 we consider the effect of additional matter 
fields in $SU_5$ representations, as discussed in 
Refs.~\ref\kmr{ C.F.~Kolda and J.~March-Russell, \prd 55 
(1997) 4252},
\ref\glr{D.~Ghilencea, M.~Lanzagorta and G.G.~Ross,
\plb 415 (1997)  253 ; \npb 511 (1998) 3\semi
G.~Amelino-Camelia, D.~Ghilencea and G.G.~Ross,
\npb 528 (1998)  35} (for earlier work see for example
Refs.~\ref\BSS{
B.~Brahmachari, U.~Sarkar and K.~Sridhar,
Mod.\ Phys.\ Lett.\ A {\bf 8}  (1993) 3349\semi
R.~Hempfling,
\plb 351 (1995)  206\semi
K.S.~Babu and J.C.~Pati,
\plb 384 (1996)  140})
and by ourselves in a previous 
paper~\ref\jjk{I.~Jack, D.R.T.~Jones and A.F.~Kord, \plb 579 (2004) 180}.
We give some further examples of the effect on the sparticle spectrum
of such matter.
Finally section~5 contains our conclusions. 

\newsec{The Soft Beta functions}

For a  general $N=1$ \sic\ gauge theory with superpotential  
\eqn\newW{
 W (\phi) = \frak{1}{2}{\mu}^{ij}\phi_i\phi_j + \frak{1}{6}Y^{ijk}
\phi_i\phi_j\phi_k,}
the standard soft \sy-breaking scalar terms are 
as follows
\eqn\newV{\eqalign{V_{\hbox{soft}} &= 
\left(\frak{1}{2}b^{ij}\phi_i\phi_j
+ \frak{1}{6}h^{ijk}\phi_i\phi_j\phi_k +\hbox{c.c.}\right)
+(m^2)^i{}_j\phi_i\phi^j,\cr}}
where we denote $\phi^i \equiv \phi_i^*$ etc. 
(For the generalisation to the case when $V_{\hbox{soft}}$ includes 
a term linear in $\phi$ 
see~\ref\jjwa{I.~Jack, D.R.T.~Jones and R.~Wild, \plb 509  131 (2001)}.) 

The complete exact results for the soft $\beta$-functions
are given by:
\eqn\allbetas{\eqalign{
\beta_M &= 2\Ocal\left[\frakk{\beta_g}{g}\right],\cr
\beta_{h}^{ijk} &= h{}^{l(jk}\gamma^{i)}{}_l -
2Y^{l(jk}\gamma_1{}^{i)}{}_l, \cr
\beta_{b}^{ij} &=   
b{}^{l(i}\gamma^{j)}{}_l-2\mu{}^{l(i}\gamma_1{}^{j)}{}_l,\cr
\left(\beta_{m^2}\right){}^i{}_j &= \Delta\gamma^i{}_j,\cr}}
where $\gamma$ is the matter multiplet anomalous dimension, and  
\eqna\Otdef$$\eqalignno{
{\cal O}  &= Mg^2\frakk{\partial}{\partial g^2}-h^{lmn}
\frakk{\partial}{\partial Y^{lmn}},
&\Otdef a\cr
(\gamma_1)^i{}_j  &= {\cal O}\gamma^i{}_j,
&\Otdef b\cr
\Delta &= 2\Ocal\Ocal^* +2MM^* g^2{\partial
\over{\partial g^2}}
+\left[\Ytil^{lmn}{\partial\over{\partial Y^{lmn}}}
+ \hbox{c.c.}\right]
+X{\partial\over{\partial g}}.&\Otdef c\cr
}$$
Here $M$ is the gaugino mass and  
$\Ytil^{ijk} = (m^2)^i{}_lY^{jkl} +  (m^2)^j{}_lY^{ikl} + (m^2)^k{}_lY^{ijl}.$
Eq.~\allbetas{}\ holds in a class of renormalisation schemes that includes 
the $\DREDp$-one\ref\jjmvy
{I.~Jack, D.R.T~Jones, S.P.~Martin, M.T.~Vaughn and Y.~Yamada,
Phys.\ Rev.\ D {\bf 50}, 5481 (1994)
}, which we will use throughout.

Finally the $X$ function above is given 
(in the NSVZ scheme
\ref\jnsvz{D.R.T.~Jones, \plb123 (1983)  45 \semi 
V.~Novikov et al, \npb229 (1983) 381  \semi
V.~Novikov et al, \plb166 (1986) 329 \semi
M.~Shifman and A.~Vainstein, \npb277 (1986)  456}) by 
\eqn\exX{
X^{\NSVZ}=-2{g^3\over{16\pi^2}}
{S\over{\left[1-2g^2 C(G)(16\pi^2)^{-1}\right]}}}
where
\eqn\Awc{
S =  r^{-1}\tr [m^2C(R)] -MM^* C(G),}
$C(R),C(G)$ being the quadratic Casimirs for the matter and adjoint 
representations respectively. 
There is no corresponding exact form 
for $X$ in  the $\DREDp$ scheme\jjmvy;
we will require the leading and sub-leading contributions, 
which are given by\ref\jjpb{I.~Jack, D.R.T.~Jones and A.~Pickering,
\plb 432 (1998) 114}:
\eqn\exXp{
\lf X^{\DREDp(1)}=-2g^3S}
and 
\eqn\exXp{
\llf X^{\DREDp(2)}= (2r)^{-1}g^3\tr [ W C(R)]
-4g^5C(G)S-2g^5C(G)QMM^*,}
where 
\eqn\Wdef{\eqalign{
W^j{}_i&={1\over2}Y_{ipq}Y^{pqn}(m^2)^j{}_n+{1\over2}Y^{jpq}Y_{pqn}(m^2)^n{}_i
+2Y_{ipq}Y^{jpr}(m^2)^q{}_r\cr &\quad
+h_{ipq}h^{jpq}-8g^2MM^*C(R)^j{}_i.\cr}}
and $Q = T(R) - 3C(G)$, and $rT(R) = \tr \left[C(R)\right]$, $r$ being 
the number of group generators. 

We now present the results for the gauge $\beta$-functions 
and anomalous dimensions. These results are valid in the 
$\DREDp$ scheme\jjmvy\  (or indeed the 
$\DRED$ one\ref\scjn{W.~Siegel,
\plb 84  (1979) 193\semi
D.M.~Capper, D.R.T.~Jones and P.~van Nieuwenhuizen,
\npb 167 (1980) 479}, which differs from $\DREDp$ only when we come to the 
soft $\beta$-functions). The MSSM superpotential is: 
\eqn\aaa{W = H_2 Q Y_t t^c  + H_1 Q Y_b b^c + H_1 L Y_{\tau} \tau^c + 
\mu H_1 H_2}
where $Y_t$, $Y_b$, $Y_{\tau}$ are $n_g\times n_g$ Yukawa matrices
\foot{$Y_{t,b,\tau}$ here are the {\it transposes\/} of the Yukawa 
matrices used in Ref.\fjja.} ,  
and we define 
\eqn\aaaa{T = Y_t^{\phantom {\dagger}}Y_t^{\dagger}, 
B = Y_b^{\phantom {\dagger}}Y_b^{\dagger}, 
E = Y_{\tau}^{\phantom {\dagger}}Y_{\tau}^{\dagger}, 
\Ttil = Y_t^{\dagger}Y_t^{\phantom {\dagger}}, 
\Btil = Y_b^{\dagger}Y_b^{\phantom {\dagger}}, 
\Etil = Y_{\tau}^{\dagger}Y_{\tau}^{\phantom {\dagger}}.}
The \smgroup\ gauge $\beta$-functions are as follows: 
\eqn\aab{
\beta_{g_i} = (\lf)^{-1}b_i g_i^3 +
(\lf)^{-2}g_i^3\left(\sum_j b_{ij}g_j^2 - a_i\right) 
+ (\lf)^{-3}\beta_{g_i}^{(3)} + \cdots}
where 
\eqn\tlfa{\eqalign{
b_1 &= \frak{1}{2}n_5 + \frak{3}{2}n_{10} +  \frak{33}{5}, 
\quad b_2 = \frak{1}{2}n_5 + \frak{3}{2}n_{10} + 1, 
\quad b_3 = \frak{1}{2}n_5 + \frak{3}{2}n_{10} - 3\cr
a_1 &= \frak{26}{5}\tr T + \frak{14}{5}\tr B + \frak{18}{5}\tr E,\quad
a_2 = 6\tr T + 6\tr B + 2\tr E,\quad
a_3 = 4\tr T + 4\tr B\cr}}
and 
\eqn\aad{
b_{ij}= \pmatrix{\frak{199}{25} + \frak{7}{30}n_5 + \frak{23}{10}n_{10} &
\frak{27}{5}+\frak{9}{10}n_5+\frak{3}{10}n_{10}
& \frak{88}{5}+ \frak{16}{15}n_5+\frak{24}{5}n_{10}
\cr 
& & \cr
\frak{9}{5}+\frak{3}{10}n_5+\frak{1}{10}n_{10}
& 
25+\frak{7}{2}n_5+\frak{21}{2}n_{10}
&
24+8n_{10}\cr
& & \cr
\frak{11}{5}+\frak{2}{15}n_5+\frak{3}{5}n_{10}
&
9+3n_{10}
& 
14+\frak{17}{3}n_5+17n_{10}
\cr
}.} 

For the anomalous dimensions of the chiral superfields we have
at one loop:

\eqn\tlfb{\eqalign{
\lf\ga^{(1)}_{t} &= 2\Ttil -\frak{8}{3}g_3^2 -\frak{8}{15}g_1^2,\cr
\lf\ga^{(1)}_{b} &= 2\Btil -\frak{8}{3}g_3^2 -\frak{2}{15}g_1^2,\cr
\lf\ga^{(1)}_{Q} &= B + T - \frak{8}{3}g_3^2 - \frak{3}{2}g_2^2
-\frak{1}{30}g_1^2,\cr
\lf\ga^{(1)}_{H_2} &= 3\tr T - \frak{3}{2}g_2^2 -\frak{3}{10}g_1^2,\cr
\lf\ga^{(1)}_{H_1} &= \tr E +3\tr B-\frak{3}{2}g_2^2
-\frak{3}{10}g_1^2,\cr
\lf\ga^{(1)}_{L} &= E - \frak{3}{2}g_2^2
-\frak{3}{10}g_1^2,\cr
\lf\ga^{(1)}_{\tau} &= 2\Etil-\frak{6}{5}g_1^2,\cr}}
and at two loops\ref\bjork{J.E.~Bj\"orkman and D.R.T.~Jones,
\npb 259 (1985) 533}:
\eqna\Keva$$\eqalignno{
\llf\gamma_t^{(2)}&=-2\Ttil^2-6(\tr T)\Ttil
-2Y_t^{\dagger} B Y_t^{\phantom {\dagger}}
+\left(6g_2^2-\frak{2}{5}g_1^2\right)\Ttil
\cr&+ (\frak{8}{15}b_1 +\frak{64}{{225}})g_1^4
+\frak{128}{{45}}g_1^2g_3^2 + (\frak{8}{3}b_3 + \frak{64}{9})g_3^4,
&\Keva a\cr
\llf\gamma_b^{(2)}&=-2\Btil^2-6(\tr B)\Btil
-2Y_b^{\dagger} T Y_b^{\phantom {\dagger}} -2(\tr E)\Btil
+\left(6g_2^2+\frak{2}{5}g_1^2\right)\Btil\cr&
+(\frak{2}{15}b_1 +\frak{4}{{225}})g_1^4
+\frak{32}{{45}}g_1^2g_3^2 +(\frak{8}{3}b_3 + \frak{64}{9})g_3^4,
&\Keva b\cr
\llf\gamma_Q^{(2)}&=-2T^2-3(\tr T)T-2B^2
-3(\tr B)B\cr&
-(\tr E)B +g_1^2(\frak{4}{5}T+\frak{2}{5}B)
+\frak{1}{{10}}g_1^2g_2^2
+8g_3^2g_2^2+\frak{8}{{45}}g_1^2g_3^2\cr&
+(\frak{8}{3}b_3 + \frak{64}{9})g_3^4
+(\frak{3}{2}b_2 + \frak{9}{4})g_2^4
+ (\frak{1}{30}b_1 + \frak{1}{{900}})g_1^4,
&\Keva c\cr
\llf\gamma_{H_2}^{(2)}&=-9\tr T^2 -3\tr BT
+\left(16g_3^2+\frak{4}{5}g_1^2\right)\tr T
+(\frak{3}{2}b_2 + \frak{9}{4})g_2^4\cr&+\frak{9}{{10}}g_1^2g_2^2
+(\frak{3}{10}b_1 + \frak{9}{100})g_1^4,
&\Keva d\cr
\llf\gamma_{H_1}^{(2)}&=-9\tr B^2 -3\tr BT-3(\tr E^2)
+\left(16g_3^2-\frak{2}{5}g_1^2\right)\tr B
+\frak{6}{5}g_1^2\tr E\cr
&+(\frak{3}{2}b_2 + \frak{9}{4})g_2^4+\frak{9}{{10}}g_1^2g_2^2
+(\frak{3}{10}b_1 + \frak{9}{100})g_1^4,
&\Keva e\cr
\llf\gamma_L^{(2)}&=-2E^2 -E(3\tr B +\tr E -\frak{6}{5}g_1^2 )
+(\frak{3}{2}b_2 + \frak{9}{4})g_2^4
\cr&+\frak{9}{{10}}g_1^2g_2^2+(\frak{3}{10}b_1 + \frak{9}{100})g_1^4,
&\Keva f\cr
\llf\gamma_{\tau}^{(2)}&=-2\Etil^2-\Etil(6\tr B +2\tr E
-6g_2^2+\frak{6}{5}g_1^2) +(\frak{6}{5}b_1 + \frak{36}{25})g_1^4.
&\Keva g\cr}$$

The three-loop gauge $\beta$-function terms are 
(henceforth we suppress $(\lf)^{-L}$ factors).  
\eqna\eva$$\eqalignno{
\beta_{g_1}^{(3)}=&g_1^3\Bigl[\frak{84}{5}\tr T^2+18(\tr T)^2 
+\frak{54}{5}\tr B^2 +\frak{36}{5}(\tr B)^2+\frak{58}{5}\tr TB
+\frak{54}{5}\tr E^2 +\frak{24}{5}(\tr E)^2\cr&+\frak{84}{5}\tr E\tr B
-\left(\frak{169}{75}g_1^2+\frak{87}{5}g_2^2 +
\frak{352}{15}g_3^2\right)\tr T -\left(\frak{49}{75}g_1^2 +
\frak{33}{5}g_2^2 +\frak{256}{15}g_3^2\right)\tr B\cr&
-(\frak{81}{25}g_1^2 + \frak{63}{5}g_2^2)\tr E
+(\frak{484}{15}-\frak{4}{5}n_5^2-(\frak{506}{45}+6 n_{10})n_5
-\frak{154}{5}n_{10}-\frak{54}{5}n_{10}^2)g_3^4\cr&
-[(\frak{24}{5}+\frak{8}{5}n_{10})g_2^2
+(\frak{64}{225}n_5+\frak{344}{75}n_{10}+\frak{1096}{75})g_1^2]g_3^2\cr&
-(\frak{27}{40}n_5^2+(\frak{27}{4}+\frak{9}{4}n_{10})n_5
+\frak{81}{5}+\frak{261}{20}n_{10}
+\frak{27}{40}n_{10}^2)g_2^4
-(\frak{1}{50}n_{10}+\frak{27}{50}n_5+\frak{69}{25})g_1^2g_2^2\cr&
-(\frak{7}{40}n_5^2+(\frak{7507}{900}+\frak{9}{4}n_{10})n_5
+\frak{12859}{300}n_{10}
+\frak{207}{40}n_{10}^2+\frak{32117}{375})g_1^4
\Bigr],
& \eva a\cr
\beta_{g_2}^{(3)}=&g_2^3\Bigl[24(\tr T^2+\tr B^2)
+18[(\tr T)^2 +(\tr B)^2] +12\tr BT+12\tr B\tr E +8\tr E^2\cr&+2(\tr
E)^2-(32g_3^2+33g_2^2)(\tr T+\tr B) 
-g_1^2(\frak{29}{5}\tr T+\frak{11}{5}\tr B)
-\left(11g_2^2+\frak{21}{5}g_1^2\right)\tr E\cr&
-[(6n_{10}+18)n_5+\frak{118}{3}n_{10}+18n_{10}^2-44]g_3^4
+[(8n_{10}+24)g_2^2
-(\frak{8}{15}n_{10}+\frak{8}{5})g_1^2]g_3^2\cr&
-(\frak{13}{8}n_5^2
+(\frak{33}{4}+\frak{39}{4}n_{10})n_5+\frak{99}{4}n_{10}
+\frak{117}{8}n_{10}^2 -35)g_2^4
+(\frak{1}{10}n_{10}+\frak{9}{5}+ \frak{3}{10}n_5)g_1^2g_2^2
\cr&
-(\frak{9}{40}n_5^2
+(\frak{3}{4}n_{10}+\frak{441}{100})n_5+\frak{9}{40}n_{10}^2
+\frak{457}{25}+\frak{1513}{300}n_{10})g_1^4
\Bigr],
&\eva b\cr
\beta_{g_3}^{(3)}=&g_3^3\Bigl[18(\tr T)^2+12\tr T^2 +8\tr BT +18(\tr
B)^2+12\tr B^2 +6\tr E\tr B\cr& -(\frak{104}{3}g_3^2+12g_2^2)(\tr T+\tr
B) -g_1^2 (\frak{44}{15}\tr T  + \frak{32}{15}\tr B)\cr&
+(\frak{347}{3}+\frak{215}{3}n_{10}-\frak{11}{4}n_5^2
+(\frak{215}{9}-\frak{33}{2}n_{10})n_5
-\frak{99}{4}n_{10}^2)g_3^4\cr&
+[(2n_{10}+6)g_2^2
+(\frak{2}{5}n_{10}+\frak{4}{45}n_5+\frak{22}{15})g_1^2]g_3^2\cr&
-[(\frak{27}{4}+\frak{9}{4}n_{10})n_5+\frak{27}{4}n_{10}^2
+\frak{117}{4}n_{10}+27]g_2^4
-(\frak{1}{5}n_{10}+\frak{3}{5})g_1^2g_2^2\cr&-(\frak{1}{10}n_5^2
+(\frak{2689}{900}+\frak{3}{4}n_{10})n_5
+\frak{27}{20}n_{10}^2+\frak{1702}{75}
+\frak{3353}{300}n_{10})g_1^4
\Bigr].
&\eva c
\cr}
$$

The three-loop results for the anomalous dimensions 
are as follows:

\eqn\newV{\eqalign{ \gamma_Q^{(3)} &= k(T^3+B^3) + 4BTB   + 4TBT  +
6T^2\tr T + B^2( 6\tr B + 2\tr E )\cr& 
+ B[ 6\tr(TB) - 9(\tr B)^2 - 6\tr B\tr E +18\tr(B^2)
- (\tr E)^2 + 6\tr(E^2) ]\cr&  -9T(\tr T)^2 +18T\tr(T^2) +6T\tr(TB)     
+  g_1^2[T^2(\frak{11}{3}- k )+B^2(\frak{7}{15}-\frak{1}{5}k)]\cr&  
+ [(3k-3)g_2^2 +\frak{64}{3}g_3^2](T^2 +B^2) 
+[( 2 -\frak{4}{5}k )g_1^2+
18g_2^2 + (8k - 8 )g_3^2]T\tr T\cr&
+g_1^2 B
[(\frak{16}{5} -\frak{4}{5}k)\tr B  +(\frak{2}{5}k-\frak{8}{5})\tr E]
 + g_2^2B( 18\tr B + 6\tr E ) \cr &  
 + 8g_3^2B[(k -1) \tr B + \tr E]
+ g_1^4T( \frak{143}{900}k -\frak{3767}{300}-\frak{51}{20}n_{10}
-\frak{17}{20}n_5 ) \cr & 
+g_1^4B( \frak{7}{180}k -\frak{633}{100}-\frak{27}{20}n_{10}
-\frak{9}{20}n_5 )   - (\frak{13}{30}\tr T +\frak{7}{30}\tr B
+\frak{3}{10}\tr E )g_1^4 \cr &  
+(\frak{3}{2}k-\frak{59}{10} )g_1^2g_2^2T 
+ (\frak{3}{10}k-\frak{41}{10} )g_1^2g_2^2B
 + g_1^2g_3^2T(\frak{64}{45}k -\frak{68}{5} ) +
g_1^2g_3^2B(\frak{64}{45}k  -\frak{76}{15})  \cr&
 - g_2^4[(T+B)( \frak{45}{4}+\frak{21}{4}k
+\frak{27}{4}n_{10} +\frak{9}{4}n_5 )
 +\frak{45}{2}(\tr T +\tr B) +\frak{15}{2}\tr E ] \cr& 
-4g_2^2g_3^2(T+B)+ 
g_3^4[(T+B)(\frak{8}{3}-\frak{136}{9}k -12n_{10} - 4n_5)
-\frak{80}{3}(\tr T +\tr B)]\cr&
+(\frak{25}{4}-\frak{3}{40}kn_{10} -\frak{9}{40}kn_5 -\frak{27}{20}k +
\frak{3}{10}n_{10} +\frak{11}{10}n_5 )g_1^2g_2^4  
-\frak{8}{5}g_1^2g_2^2g_3^2\cr& 
+ g_1^6(\frak{28457}{13500}-k(
\frak{23}{600}n_{10} +\frak{7}{1800}n_5 +\frak{199}{1500}) + 
(\frak{1}{20}n_5 +\frak{17}{20} +\frak{3}{40}n_{10})n_{10} +
\frak{43}{180}n_5 +\frak{1}{120}n_5^2 )\cr& +
g_1^4g_2^2(\frak{11}{100}-\frak{1}{200}kn_{10}
-\frak{3}{200}kn_5 -\frak{9}{100}k -\frak{1}{20}n_{10} +\frak{1}{20}n_5
)\cr& + g_1^4g_3^2(\frak{194}{225} -\frak{2}{25}kn_{10}
-\frak{4}{225}kn_5 -\frak{22}{75}k +\frak{4}{15}n_{10} +\frak{2}{45}n_5
)\cr& +
g_1^2g_3^4(\frak{608}{45}-\frak{4}{5}kn_{10}
-\frak{8}{45}kn_5 -\frak{44}{15}k +\frak{58}{15}n_{10}
+\frak{38}{45}n_5 ) \cr& 
+g_2^4g_3^2( 50 - 6kn_{10} -18k + 24n_{10} - 2n_5 )
+g_2^2g_3^4( 8 - 4kn_{10} -12k + 14n_{10} -2n_5 )\cr& 
+g_2^6(\frak{345}{4}+\frak{45}{8}kn_{10} +\frak{15}{8}kn_5 +
\frak{105}{4}k +\frak{9}{4}n_{10}n_5 +\frak{81}{2}n_{10}
+\frak{27}{8}n_{10}^2 +\frak{27}{2}n_5 +\frak{3}{8}n_5^2 )\cr& 
+ g_3^6(\frak{2720}{27}+k(
\frak{40}{3}n_{10} +\frak{40}{9}n_5 +\frak{160}{3}) + n_{10}(4n_5 
+\frak{236}{3} + 6n_{10}) +\frak{236}{9}n_5 +\frak{2}{3}n_5^2 )
}}
\eqn\newV{\eqalign{ \gamma_L^{(3)} &= k E^3 + E^2( 6\tr B +
2\tr E )\cr& 
+ E[6\tr(TB) - 9(\tr B)^2 - 6\tr B\tr E +18\tr(B^2) -
(\tr E)^2 + 6\tr(E^2)]\cr& + g_1^2E^2( 9 -\frak{9}{5}k ) 
+(8-2k) g_1^2E\tr B 
+ (3k-3 )g_2^2E^2 + g_2^2E( 18\tr B + 6\tr E )
\cr&+ (8k  - 32)g_3^2E\tr B
+ g_1^4E( -\frak{549}{20} +\frak{27}{100}k
-\frak{99}{20}n_{10} -\frak{33}{20}n_5 )  
\cr& + g_1^4( -
\frak{39}{10}\tr T -\frak{21}{10}\tr B -\frak{27}{10}\tr E)\cr& +
g_1^2g_2^2E( -\frak{81}{10}+\frak{27}{10}k ) 
 + g_2^4E(
-\frak{45}{4}-\frak{21}{4}k -\frak{27}{4}n_{10} -\frak{9}{4}n_5 ) 
\cr&
+ g_2^4( -\frak{45}{2}\tr T
-\frak{45}{2}\tr B -\frak{15}{2}\tr E )+ \Xi 
}}

\eqn\newV{\eqalign{ \gamma_t^{(3)} &= ( 6 + 2k )\Ttil^3 +
6\Ttil^2\tr T  -2Y_t^{\dagger}BTY_t -2Y_t^{\dagger}TBY_t
+6Y_t^{\dagger}B^2Y_t\cr& 
+ \Ttil( 36\tr(T^2) +12\tr(T B) -18(\tr T)^2  ) +
Y_t^{\dagger}BY_t( - 6\tr T +12\tr B + 4\tr E )\cr& 
 + g_1^2\left[\Ttil^2( -\frak{1}{3}+ k )
+ 7(1 +\frak{k}{5})\Ttil\tr (T) +
Y_t^{\dagger}BY_t(\frak{19}{15}+\frak{3}{5}k )\right]\cr& 
+g_2^2\left[( 9 - 3k )\Ttil^2 +( 27 -9k )\Ttil\tr T 
+( 9 - 3k ) Y_t^{\dagger}BY_t\right]\cr&
+ g_3^2\left[16(k-1) \Ttil\tr T
+ \frak{64}{3}(\Ttil^2+Y_t^{\dagger}BY_t)\right]\cr&
+g_1^4\left[\Ttil( - \frak{799}{50}- \frak{247}{450}k
-\frak{18}{5}n_{10} - \frak{6}{5}n_5 )  -
\frak{104}{15}\tr T - \frak{56}{15}\tr B - \frak{24}{5}\tr E\right]
\cr&+ g_1^2g_3^2\Ttil(
-\frak{8}{15}- \frak{112}{45}k )  +
g_1^2g_2^2\Ttil( - \frak{67}{5}+ \frak{13}{5}k )\cr& 
 +
g_2^4\Ttil(
 - \frak{87}{2}-\frak{3}{2}k -18n_{10} - 6n_5 ) +
g_2^2g_3^2\Ttil( - 88 +16k ) \cr& 
 +
g_3^4\left[\Ttil(\frak{16}{3}- \frak{272}{9}k - 24n_{10} - 8n_5 ) 
-\frak{80}{3}\tr T -\frak{80}{3}\tr B\right]\cr& 
+ g_1^2g_3^4( -
\frak{172}{45}-\frak{4}{5}kn_{10} -\frak{8}{45}kn_5 -
\frak{44}{15}k + \frak{28}{15} n_{10} + \frak{8}{45}n_5 )\cr&
 +
g_1^4g_2^2( \frak{36}{5}-\frak{2}{25}kn_{10} -
\frak{6}{25}kn_5 - \frak{36}{25}k + \frak{2}{5}n_{10} +\frak{6}{5} n_5
)\cr& + g_1^4g_3^2( \frak{2144}{225} - \frak{32}{25}kn_{10} -
\frak{64}{225}kn_5 - \frak{352}{75}k +\frak{64}{15} n_{10} +
\frak{32}{45}n_5 )\cr&
+ g_3^6(\frak{2720}{27}+
\frak{40}{3}kn_{10} +\frak{40}{9}kn_5 +\frak{160}{3}k + 4n_{10}n_5 +
\frak{236}{3}n_{10} + 6n_{10}^2 +\frak{236}{9}n_5 +\frak{2}{3}n_5^2)
\cr&
+g_1^6(
\frak{106868}{3375} - k(\frak{46}{75}n_{10} + \frak{14}{225}n_5 +
\frak{796}{375}) +n_{10}(\frak{4}{5}n_5 + \frak{66}{5}
+\frak{6}{5}n_{10}) +\frak{166}{45}n_5 +\frak{2}{15}n_5^2 )\cr&
+g_2^2g_3^4( 60 -
4kn_{10} -12k +20n_{10})}}

\eqn\gahonec{\eqalign{ 
\gamma_{H_1}^{(3)} &= (k+1)\left[3\tr(B^3)+\tr(E^3)\right] 
+ 9\tr(T^2B)+ 18\tr T\tr(TB) \cr&
+ 6(\tr E+3\tr B)\left[\tr(E^2)+3\tr(B^2)\right]  
+ (24-8k)g_3^2 [\tr(TB) + 3\tr(B^2)]\cr&
+ g_2^2[ 18\tr(TB)
+ (9k+9)\tr(B^2) + (3k+3)\tr(E^2)]\cr&
+ g_1^2[
-\frak{12}{5}\tr(TB) +\frak{7}{5}k\tr(TB) + 3\tr(B^2)
+\frak{9}{5}k\tr(B^2) + 9\tr(E^2) -\frak{9}{5}k\tr( E^2) ]\cr& 
+g_1^4[ -\frak{39}{10}\tr T - \tr B(\frak{175}{12} +
\frak{77}{300}k +\frak{57}{20}n_{10} +\frak{19}{20}n_5) 
+\tr E(\frak{27}{100}k-\frak{603}{20}\cr&  - \frak{99}{20}n_{10} -
\frak{33}{20}n_5) ] +
g_1^2g_2^2( -\frak{3}{10}\tr B -\frak{3}{2}k\tr B
-\frak{81}{10}\tr E +\frak{27}{10}k\tr E )\cr& 
+ \tr B[g_1^2g_3^2(
-\frak{284}{15} + \frak{56}{15}k ) 
+ g_2^2g_3^2( -132+ 24k )]
+g_2^4[-\frak{45}{2}\tr T\cr&-\tr B(\frak{225}{4} + \frak{63}{4}k
+\frak{81}{4}n_{10} +\frak{27}{4}n_5) - \tr E(\frak{75}{4}
+\frak{21}{4}k +\frak{27}{4}n_{10} +\frak{9}{4}n_5)]
\cr&  -g_3^4\tr B(
\frak{160}{3} +\frak{8}{3}k + 48n_{10} +16n_5 )+ \Xi 
}}

\eqn\newV{\eqalign{ \gamma_b^{(3)} &= (2k+6)\Btil^3 
+6Y_b^{\dagger}T^2Y_b -2Y_b^{\dagger}BTY_b 
+ Y_b^{\dagger}TY_b( 12\tr T - 6\tr B - 2\tr E
)\cr& -2Y_b^{\dagger}TBY_b  + \Btil^2(
6\tr B + 2\tr E )\cr& + \Btil[12\tr(TB) - 18(\tr B)^2
-12\tr B\tr E + 36\tr(B^2) - 2(\tr E)^2 +12\tr(E^2 )]\cr& +
g_1^2Y_b^{\dagger}TY_b( -\frak{29}{15}+\frak{3}{5}k ) +
g_1^2\Btil^2( -\frak{1}{3}+\frak{1}{5}k ) 
+ g_1^2\Btil[(7 - k)\tr B + (k-3)\tr E ]\cr&  
+g_2^2Y_b^{\dagger}TY_b( 9 - 3k ) +
g_2^2\Btil^2( 9 - 3k ) 
+ g_2^2\Btil[( 27-9k)\tr B+(9-3k)\tr E ]\cr&
 + \frak{64}{3}g_3^2Y_b^{\dagger}TY_b+ \frak{64}{3}g_3^2\Btil^2 + 
g_3^2\Btil[(16k-16)\tr B+16\tr E]\cr&
+g_1^4\left[\Btil( -\frak{337}{30}-\frak{7}{450}k -\frak{12}{5}n_{10}
-\frak{4}{5}n_5 )-\frak{26}{15}\tr T
-\frak{14}{15}\tr B -\frak{6}{5}\tr E\right] \cr& 
+ g_1^2g_2^2\Btil(
-\frak{43}{5}+\frak{7}{5}k ) + g_1^2g_3^2\Btil(
-\frak{24}{5}+\frak{16}{9}k ) +
g_2^2g_3^2\Btil( - 88 +16k )\cr&+
g_2^4\Btil( -\frak{87}{2}-\frak{3}{2}k -18n_{10} - 6n_5 )\cr&
 + g_3^4\left[\Btil(\frak{16}{3}-\frak{272}{9}k -
24n_{10} - 8n_5 )
 -\frak{80}{3}\tr T
-\frak{80}{3}\tr B\right] \cr&
+g_3^6(\frak{2720}{27}+\frak{40}{3}kn_{10} +\frak{40}{9}kn_5
+\frak{160}{3}k + 4n_{10}n_5 + \frak{236}{3}n_{10} + 6n_{10}^2 +
\frak{236}{9}n_5 +\frak{2}{3}n_5^2 )\cr&
+g_1^6(\frak{5629}{675}-k(\frak{23}{150}n_{10}
+\frak{7}{450}n_5 +\frak{199}{375}) + n_{10}(\frak{1}{5}n_5 +
\frak{169}{50} +\frak{3}{10}n_{10}) +\frak{427}{450}n_5
+\frak{1}{30}n_5^2 )\cr& +
g_1^4g_2^2(\frak{9}{5}-\frak{1}{50}kn_{10} -\frak{3}{50}kn_5
-\frak{9}{25}k +\frak{1}{10}n_{10} +\frak{3}{10} n_5 )\cr& +
g_1^4g_3^2( \frak{728}{225} -\frak{8}{25}kn_{10}
-\frak{16}{225}kn_5 - \frak{88}{75}k +\frak{16}{15}n_{10}
+\frak{8}{45}n_5 )\cr&
+ g_1^2g_3^4(\frak{452}{45}
-\frak{4}{5}kn_{10} -\frak{8}{45}kn_5 -\frak{44}{15}k +
\frak{52}{15}n_{10} + \frak{32}{45}n_5 )\cr&
+ g_2^2g_3^4( 60 -
4kn_{10} -12k +20n_{10})}}

\eqn\newV{\eqalign{ \gamma_{H_2}^{(3)} &=  
 54\tr T\tr(T^2) + 3(k+1)\tr(T^3) 
+18\tr B\tr(TB) + 9\tr(BTB) + 6\tr E\tr(TB)
\cr&
+g_1^2(
(\frak{57}{5}-\frak{3}{5}k)\tr(T^2)  + (\frak{6}{5}
+\frak{1}{5}k)\tr(TB) )
+ g_2^2[(9+ 9k)\tr(T^2)  +18\tr(TB)]\cr&
+ g_3^2[(72 - 24k)\tr(T^2) + (24-8k)\tr(TB)] \cr&  
-g_1^4\left[( \frak{2123}{60} +
\frak{13}{60}k +\frak{43}{20}n_5 +\frak{129}{20}n_{10})\tr T  
+\frak{21}{10}\tr B +\frak{27}{10}\tr E\right]\cr&  
+[g_1^2g_2^2( \frak{21}{10}k-\frak{57}{10}  ) +
g_1^2g_3^2(\frak{104}{15}k -\frak{124}{3}  ) 
+ g_2^2g_3^2( 24k - 132  )]\tr T\cr&  
+ g_2^4( \tr T (-\frak{225}{4} -
\frak{63}{4}k -\frak{81}{4}n_{10} -\frak{27}{4}n_5) -
\frak{45}{2}\tr B -\frak{15}{2}\tr E )\cr&
-g_3^4(\frak{160}{3}\tr T +\frak{8}{3}k\tr T + 48n_{10}\tr T +
16n_5\tr T)+\Xi}}

\eqn\newV{\eqalign{ \gamma_{\tau}^{(3)} &= (6 + 2k)\Etil^3 +
( 6\tr B + 2\tr E )\Etil^2\cr& + \Etil( 12\tr(TB) -18(\tr B)^2 -
12\tr B\tr E + 36\tr(B^2) - 2(\tr E)^2 +12\tr(E^2 ) )\cr& +
g_1^2\Etil^2(\frak{9}{5}+\frak{9}{5}k ) 
+ g_2^2\Etil^2( 9-3k ) + g_1^2\Etil(
(\frak{107}{5} +\frak{7}{5}k)\tr B 
+(\frak{9}{5} +\frak{9}{5}k)\tr E)\cr& 
+g_3^2\Etil\tr B(16k - 64  )
+ g_2^2\Etil(\tr B( 27 - 9k) + \tr E(9 - 3k) )\cr&
 + g_1^4\Etil( - \frak{1503}{50}
-\frak{27}{10}k - \frak{36}{5}n_{10} -\frak{12}{5}n_5 )
+g_1^2g_2^2\Etil( - 27 +\frak{27}{5}k )\cr& 
+ g_1^4( -
\frak{78}{5}\tr T - \frak{42}{5}\tr B - \frak{54}{5}\tr E ) 
+ g_2^4\Etil( -\frak{87}{2}-\frak{3}{2}k -18n_{10} - 6n_5 )
\cr&
+g_1^6( \frak{7899}{125} - \frak{69}{50}kn_{10}
-\frak{7}{50}kn_5 - \frak{597}{125}k +\frak{9}{5}n_{10} n_5 +
\frak{57}{2}n_{10} +\frak{27}{10}n_{10}^2 + \frak{79}{10}n_5
+\frak{3}{10}n_5^2 )\cr& +
g_1^4g_2^2(\frak{81}{5}-\frak{9}{50}kn_{10} -
\frak{27}{50}kn_5 -\frak{81}{25}k +\frak{9}{10}n_{10} +\frak{27}{10}n_5 )\cr& +
g_1^4g_3^2(\frak{264}{5}- \frak{72}{25}kn_{10} -\frak{16}{25}kn_5 -
\frak{264}{25}k + \frak{72}{5}n_{10} + \frak{16}{5}n_5 )\cr&  
}}
where $k = 6\zeta(3)$, and 
\eqn\Xidef{\eqalign{\Xi &= 
g_2^6(\frak{345}{4}+k(\frak{45}{8}n_{10} +\frak{15}{8}n_5 +
\frak{105}{4}) +n_{10}(\frak{9}{4}n_5 +\frak{81}{2} +\frak{27}{8}n_{10})
+\frak{27}{2}n_5 +\frak{3}{8}n_5^2 )\cr&  
+ g_1^6(
\frak{1839}{100}-k(\frak{69}{200}n_{10} +\frak{7}{200}n_5 +\frak{597}{500})
+n_{10}(\frak{9}{20}n_5 +\frak{753}{100} +\frak{27}{40}n_{10}) +
\frak{211}{100}n_5 +\frak{3}{40}n_5^2 )\cr& 
+ g_1^4g_2^2(\frak{27}{100}
-\frak{9}{200}kn_{10} -\frak{27}{200}kn_5 -\frak{81}{100}k
-\frak{9}{20}n_{10} +\frak{9}{20}n_5 )\cr& + g_1^4g_3^2(\frak{66}{5}-
\frak{18}{25}kn_{10} -\frak{4}{25}kn_5 -\frak{66}{25}k +\frak{18}{5}n_{10}
+\frak{4}{5}n_5)\cr&
+g_1^2g_2^4(\frak{9}{4}-\frak{3}{40}kn_{10} -\frak{9}{40}kn_5
-\frak{27}{20}k -\frak{3}{10}n_{10} +\frak{9}{10}n_5 )\cr&
+ g_2^4g_3^2( 90 - 6kn_{10} -18k + 30n_{10}).}}

In terms of the anomalous dimensions, the Yukawa $\beta$-functions are:
\eqn\tlfb{
\beta_{Y_t} = \gamma_Q Y_t + Y_t ( \gamma_t + \gamma_{H_2})
,\quad
\beta_{Y_b} = \gamma_Q Y_b + Y_b ( \gamma_b + \gamma_{H_1})
,\quad
\beta_{Y_{\tau}} = \gamma_L Y_{\tau}
+ Y_{\tau} ( \gamma_{\tau} + \gamma_{H_1}),
}
and the $\beta$-function for the Higgs $\mu$-term is
\eqn\betamu{\beta_{\mu} = \mu(\gamma_{H_2}+\gamma_{H_1}).}
We will also require the anomalous dimensions of the  constituents of
the extra $5$ and $10$ representations, which   are easily obtained by
setting  $T = B = E = 0$, except retaining terms that contain $T,B,E$
{\it only\/}  inside traces; such terms occur for the  first time at
three loops.

From the above expressions for $\beta_{g_i}$ and $\gamma$ we have 
calculated the three-loop soft $\beta$-functions using 
Eq.~\allbetas\ and  FORM. The 
resulting expressions are very unwieldy; as an example 
we give the one, two and three-loop results for $\beta_{m^2_{Q_t}}$, 
in the approximation that we retain only $g_3^2$ and the 
top quark Yukawa coupling $\lambda_t$ (in what follows 
we denote the third generation squarks as $Q_t, t^c, b^c$,
and the first or second generation squarks as 
$Q_u, u^c, d^c$):

\eqna\bqthree$$\eqalignno{
\beta_{m^2_{Q_t}}^{(1)} &= 
2 \lambda_t^2  (\Sigma_t +A_t^2)
-8(\frak{1}{60}g_1^2 M_1^2+\frak{3}{4} g_2^2  M_2^2+\frak{4}{3} g_3^2 M_3^2) 
& \bqthree a\cr
\beta_{m^2_{Q_t}}^{(2)} &= -20\lambda_t^4(\Sigma_t+2A_t^2)
+16g_3^4M_3^2(n_5+3n_{10}
-\frak{8}{3})\cr&
+\frak{16}{3}g_3^4 (2m^2_{Q_t}+m^2_{t^c}+m^2_{b^c}
+ (n_{10}+2)(m^2_{u^c}+2m^2_{Q_u})
+(n_5+2)m^2_{d^c}) & \bqthree b\cr
\beta_{m^2_{Q_t}}^{(3)} &=
[(1280k+\frak{20512}{9}+16n_5^2 +(\frak{6224}{9}+
\frak{320}{3}k)(n_5+3n_{10})\cr& + 96n_{10}n_5 +144n_{10}^2)M_3^2
+(\frak{320}{9}-\frak{16}{3}(n_5+3n_{10}))
(m^2_{t^c}+m^2_{b^c}+2m^2_{Q_t})\cr& +(2m^2_{Q_u}+m^2_{u^c})(\frak{640}{9}
-\frak{32}{3}n_5 + \frak{32}{9}n_{10}
     -\frak{16}{3}n_5n_{10}-16n_{10}^2)\cr&
+m^2_{d^c}(\frak{640}{9} +\frak{224}{9}n_5 -32n_{10} 
     -16n_5n_{10}-\frak{16}{3}n_{5}^2)] g_3^6\cr& 
-[(288+\frak{544}{3}k + 48(n_5+3n_{10}))M_3^2
-(192+\frak{1088}{9}k +32(n_5+3n_{10}))A_tM_3\cr&
+(\frak{272}{9}k+\frak{176}{3}+8(n_5+3n_{10}))(\Sigma_t+A_t^2)]\lambda_t^2g_3^4\cr&
+(\frak{160}{3}+32k)\left[M_3^2-2A_tM_3
+\Sigma_t
+2A_t^2\right]\lambda_t^4g_3^2\cr
&+(6k+90)(\Sigma_t+3A_t^2)\lambda_t^6, & \bqthree c}$$
where $\Sigma_t = m^2_{Q_t}+m^2_{2}+m^2_{t^c}$.
For this special case, and also with $n_5 = n_{10} = 0$,  the three-loop
result, Eq.~\bqthree{c}, was  given in  
Ref.~\ref\KazakovBT{D.I.~Kazakov, hep-ph/0208200}, 
except that in the corresponding expressions in this 
reference the squark  masses of different generations are not
clearly distinguished (as they must  be since the third generation evolves
differently from the other two).
Complete results for the three-loop $\beta$-functions including 
all three gauge couplings and  $n_g\times n_g$ Yukawa matrices
are available at Ref.~\website. 

In our analysis we do  include ``tadpole'' contributions,
corresponding to renormalisation of the Fayet-Iliopoulos (FI) $D$-term 
at one and two loops. 
These contributions are not expressible exactly in terms of 
$\beta_{g_i}, \gamma$; for a discussion, and three-loop results for the 
MSSM,  see 
Ref.~\ref\jjp{I.~Jack and D.R.T.~Jones, \plb473 (2000) 102\semi
I.~Jack, D.R.T.~Jones and S.~Parsons, \prd62 (2000) 125022\semi
I.~Jack and D.R.T.~Jones, \prd63 (2001) 075010}. For 
universal boundary conditions, the FI term is very small at 
low energies if it is zero at gauge unification; including the three-loop
(FI) effects would have a negligible effect on our results.

\newsec{The Snowmass Benchmark Points}

In this section we examine the effect of the three-loop corrections 
on the standard running analysis, that is for $n_5 = n_{10} = 0$. 
We will focus on the standard treatment 
with universal boundary conditions at gauge unification, often 
termed CMSSM or MSUGRA. Thus we assume that at $M_X$ we 
have universal soft scalar masses ($m_0$), 
gaugino masses\foot{except for the SPS6 point.
The SPS6 point corresponds to non-unified gaugino masses, 
$M_1 = 480\GeV, M_2 = M_3 = 300\GeV$.}
($m_{\frak{1}{2}}$)
and $A$-parameters ($A$), and work in the third-generation-only 
Yukawa coupling approximation.  This is for ease of comparison with 
existing results rather than because we find the scenario particularly 
compelling. We will present results for the set of 
MSUGRA Snowmass Benchmark Points shown in Table~1:

\vskip3em   
\vbox{
\begintable
{\rm Point} |  \tan\beta  |  m_{\half}  |  m_0  |  A  |  \hbox{sign} \mu  \cr
{\rm SPS}1a |  10  |  250\GeV  |  100\GeV   |   -100\GeV  | +
\cr
{\rm SPS}1b |  30  |  400\GeV  |  200\GeV   |   0  | +
\cr
{\rm SPS}2 |  10  | 300\GeV  |  1450\GeV  | 0 | +
\cr
{\rm SPS}3 |  10  | 400\GeV  |  90\GeV  | 0 | +
\cr
{\rm SPS}4 |  50  | 300\GeV  |  400\GeV  |  0  | +
\cr
{\rm SPS}5 |  5  | 300\GeV  |  150\GeV  |  -1\TeV  | +
\cr
{\rm SPS}6 |  10  | \hbox{see footnote}^3  |  150\GeV  | 0 | +
\endtable}
\centerline{{\it Table~1:\/} Input parameters for the SPS Benchmark Points}        
\medskip

Other input parameters are shown in Table~2:
\vskip3em
\vbox{
\begintable
 m^{\hbox{pole}}_t  | m^{\hbox{pole}}_b  
| m^{\hbox{pole}}_{\tau}
  | \alpha_3 (M_Z) |  \alpha_2 (M_Z) | \alpha_1 (M_Z)  \cr
  178\GeV | 4.9\GeV   |  1.777\GeV  |0.1172     | 0.033823    | 0.016943
\endtable}
\centerline{{\it Table~2:\/} Input parameters for the running analysis}        
\medskip

In Table~2 the input couplings $\alpha_{1\cdots 3}$  correspond to the
Standard Model $\msbar$ results; we calculate the appropriate
dimensionless coupling input  values  for the running analysis by an
iterative procedure involving the sparticle spectrum. 
We define 
gauge unification to be the scale where $\alpha_2$ and $\alpha_1$ meet;
we speed up the determination of this by (at each iteration) adjusting 
the unification scale using the solution of the one-loop $\beta$-functions 
for the gauge couplings from the previous value of the scale.
We employ one-loop radiative corrections as detailed in 
Ref.~\ref\pbmz{
D.M.~Pierce, J.A.~Bagger, K.T.~Matchev and R.J.~Zhang,
\npb 491 (1997) 3 
}
\foot{In the first line of Eq.~37 of Ref.~\pbmz, 
the first term in the square bracket should read 
$-(m_{\ttil_1}^2 
+ m_{\ttil_2}^2)B_0 ( m_{\ttil_2}, m_{\ttil_1},0)$: 
i.e. it should have a minus sign. The corresponding exact result 
in Eq.~D49 is correct, however.}; thus we 
run up from $M_Z$ using the full \sic\ $\beta$-functions.
For most particles we evaluate the pole mass at a 
renormalisation scale  equal to the pole mass 
itself, and determine this value  by iteration; the exception being the 
light CP-even Higgs, where we use a scale equal to the average squark mass.

\subsec {Benchmark point SPS 1a}

This point is a ``typical'' point in MSUGRA parameter space. 
In Table~3 we  compare  our results for a selection of sparticle 
masses (at $n_5 = n_{10} = 0$) with the spread of results taken from
Ref.~\sabine, denoted AKP (note our convention that the predominantly
right-handed top squark is $\ttil_2$).
\vskip3em
\vfil\break
\vbox{
\begintable
 {\rm mass} | 1{\rm loop}  |2{\rm loops}| 3{\rm loops} | {\rm AKP} \cr
 {\tilde g} | 628 | 613 |611 | 604-612\cr
 \ttil_1 | 594| 590| 583| 577-588\cr
 \ttil_2 | 400| 399| 391| 396-401\cr
 \util_L  |573 | 565 |557 | 565-569 \cr
\util_R | 552| 548| 539| 547-549 \cr
\btil_1 |520  |514 |507   | 514-518\cr
 \btil_2 |551  |548 | 540  | 539-548\cr
\dtil_L | 579 |571| 563 | 571-574\cr
 \dtil_R | 551 | 548|539 | 546-548\cr
\tautil_1 |212  |207 |206   | 208-211\cr
 \tautil_2 |139  |135 | 135  | 134-136\cr
\etil_L | 209 |202 |202 | 204-207\cr
 \etil_R | 147 |144|144 | 143-146\cr
\nutil_e  |192 | 186 | 185 | 186-191\cr
\nutil_{\tau}  |191 | 185 |184  | 185-191\cr 
\chi_1   | 104 | 97| 97|95.6-97.4\cr
\chi_2  |193 | 180 | 179 |181-182 \cr
\chi_3  | 351| 369 | 364 | 362-371\cr
\chi_4  | 376| 388 |384  | 381-390\cr
\chi^{\pm}_1  |193 | 179 |178 |   180-182  \cr
\chi^{\pm}_2  |376 | 388 | 384 | 380-390 \cr
h  |114 |114  | 114 | 112-115\cr
H  |392 |403  | 399 | 403-407\cr
A  |391 |403  |399  | 400-406\cr
H^{\pm}  |400|412  |408  |410-415 
\endtable}

\centerline{{\it Table~3:\/} Sparticle masses    
(in $\GeV$) for the SPS1a point}        
\medskip

We would expect our two-loop results to correspond most
closely to AKP and we see that they are indeed broadly consistent,
typically being within the range defined by the other 
programs or within a GeV of it. 
The effect of inclusion of three-loop running is never greater than 
$2\%$; note, however,  that the shift 
caused by three-loop running effects is comparable for $\util_{L}$ 
and larger for $\ttil_2, \util_{R}$ than
that produced  by two-loop running effects.

\subsec {Benchmark point SPS 1b}
This is another ``typical'' point but with a higher value of $\tan\beta$.
Our results are given in Table~4.
\vskip3em
\vbox{
\begintable
 {\rm mass} | 1{\rm loop}  |2{\rm loops}| 3{\rm loops} | {\rm AKP} \cr
 {\tilde g} | 967 | 946 |943 | 933-943 \cr
 \ttil_1 | 848| 841| 832| 836-839\cr
 \ttil_2 | 657| 656| 646| 652-661\cr
 \util_L  |891 | 878 | 868| 878-882 \cr
\util_R | 854| 849| 837| 848-850 \cr
\btil_1 |781 |773 |763  | 773-778\cr
 \btil_2 | 831 |827 | 816  | 819-828\cr
\dtil_L | 895 |882| 872 |882-885 \cr
 \dtil_R | 851 | 847|835 | 844-848 \cr
\tautil_1 |353  |347 |346   | 347-349\cr
 \tautil_2 |208  |199 | 200  | 196-202\cr
\etil_L | 348 |339 |338 | 341-342\cr
 \etil_R | 258 |254|254 | 253-256\cr
\nutil_e  |338 | 329 | 328 | 329-332\cr
\nutil_{\tau}  |328 | 318 |318  | 319-322\cr 
\chi_1       | 173 |162| 162 | 159-163\cr
\chi_2       |327 | 305| 304 |308-308 \cr
\chi_3       | 507| 532 | 526 | 521-534\cr
\chi_4       | 526| 546 |541  | 534-546\cr
\chi^{\pm}_1 |327 | 305 |  304    | 307-308  \cr
\chi^{\pm}_2  |526 | 547 | 541| 535-547\cr
h       |118 |118  | 118 | 117-119\cr
H       |528 |544  | 539 | 540-544\cr
A       |529 |545  |540 | 538-544\cr
H^{\pm}|535|551 |547 |547-551 
\endtable}

\centerline{{\it Table~4:\/} Sparticle masses    
(in $\GeV$) for the SPS1b point}        
\medskip

\subsec {Benchmark point SPS 2}
This is a ``focus point region'' point
\ref\FengZG{
J.L.~Feng, K.T.~Matchev and T.~Moroi,
\prd 61  (2000) 075005}, 
characterised by the large value of $m_0$. 
Our results are given in Table~5.
\vskip3em
\vbox{
\begintable
 {\rm mass} | 1{\rm loop}  |2{\rm loops}| 3{\rm loops} | {\rm AKP} \cr
 {\tilde g} | 835 | 816 |814 | 778-805 \cr
 \ttil_1 | 1322|1292| 1287| 1291-1318\cr
 \ttil_2 | 942|921| 913| 913-942\cr
 \util_L  |1597 | 1562 |1558| 1566-1591 \cr
\util_R | 1584| 1556| 1552| 1556-1581 \cr
\btil_1 |1303 |1273 |1268 | 1280-1309\cr
 \btil_2 | 1571 |1544 | 1540  | 1527-1568\cr
\dtil_L | 1600 |1564| 1560 |1567-1593\cr
 \dtil_R | 1584 | 1556|1553 | 1555-1580 \cr
\tautil_1 |1463  |1454 |1454   | 1455-1460\cr
 \tautil_2 |1444  |1440 | 1441  | 1439-1443\cr
\etil_L | 1468 |1459 |1459 | 1460-1465\cr
 \etil_R | 1457 |1453|1453 | 1453-1455\cr
\nutil_e  |1465 | 1456 | 1456 | 1457-1463\cr
\nutil_{\tau}  |1459 | 1450 |1450  | 1451-1457\cr 
\chi_1       | 132 |123| 123 | 121-124\cr
\chi_2       |257 | 237| 237 |240-241\cr
\chi_3       | 562| 579 | 582 | 528-596\cr
\chi_4       | 574| 589 |592  | 539-605\cr
\chi^{\pm}_1 |257 | 237 |  237    | 240-241  \cr
\chi^{\pm}_2  |574 | 590 | 592| 539-605\cr
h       |119 |119  | 119 | 117-117\cr
H       |1548 |1545 | 1546 | 1542-1555\cr
A       |1548 |1545  |1546 | 1532-1555\cr
H^{\pm}|1550|1547 |1548 |1544-1557 
\endtable}

\centerline{{\it Table~5:\/} Sparticle masses    
(in $\GeV$) for the SPS2 point}        
\medskip

\subsec {Benchmark point SPS 3}

This is a ``co-annihilation region'' point, 
its distinctive feature being a light stau not much heavier than the 
neutralino LSP. Our results are given in Table~6.
\vskip3em
\vbox{
\begintable
 {\rm mass} | 1{\rm loop}  |2{\rm loops}| 3{\rm loops} | {\rm AKP} \cr
 {\tilde g} | 964  | 943 |940 |930-940 \cr
 \ttil_1 | 851| 845|835| 836-843\cr
 \ttil_2 | 645| 644| 634| 640-650\cr
 \util_L  |872 | 860|849 | 861-863 \cr
\util_R | 835| 830| 818| 828-831 \cr
\btil_1 |794 |787 |776  | 786-793\cr
 \btil_2 |830 |826 | 814  | 816-825\cr
 \dtil_L | 876 | 864|853 |864-867 \cr
\dtil_R | 831 | 828| 816 |825-829 \cr
\tautil_1 |300  |291 |290  | 293-294\cr
 \tautil_2 |180  |173 | 173  |172-176 \cr
 \etil_L |299  | 288|288 | 291-293\cr
\etil_R | 186 | 181| 181 | 179-183\cr
\nutil_e  |287 | 277 | 276 | 277-281\cr
\nutil_{\tau}  |286 |276  |275  | 276-280 \cr 
\chi_1   | 172 | 161| 161|158-162\cr
\chi_2  |325 | 302 | 301 | 305-306\cr
\chi_3  |512 | 538 |531  | 528-540\cr
\chi_4  |533 | 554 | 548 | 543-555\cr
\chi^{\pm}_1  |324 |302  |301  |304-306 \cr
\chi^{\pm}_2  |533 |554  |548  | 542-555 \cr
h  |117 | 118 | 117 | 116-118\cr
H  |579 |597  |591  | 593-600\cr
A  |579 |597  |591  | 589-600\cr
H^{\pm}  |585 |603  | 597 |598-605 
\endtable}

\centerline{{\it Table~6:\/} Sparticle masses    
(in $\GeV$) for the SPS3 point}        
\medskip

\subsec {Benchmark point SPS4}

This is a point with large $\tan\beta$. Our results are given in Table~7.

\vskip3em
\vbox{
\begintable
 {\rm mass} | 1{\rm loop}  |2{\rm loops}| 3{\rm loops} | {\rm AKP} \cr
 {\tilde g}     | 759  | 743 | 741  | 729-738 \cr
 \ttil_1        | 705  | 700 | 693  | 693-697\cr
 \ttil_2        | 544  | 541 | 533  | 540-544\cr
 \util_L        | 777  | 764 | 757  | 766-772 \cr
\util_R         | 755  | 747 | 739  | 747-751 \cr
\btil_1         | 624  | 619 | 611  | 614-619\cr
 \btil_2        |693   | 690 | 683  | 679-692\cr
 \dtil_L        | 782  | 769 | 761  | 770-776 \cr
\dtil_R         | 753  | 746 | 738  | 746-749 \cr
\tautil_1       | 423  | 420 | 420  | 414-421\cr
 \tautil_2      |272   | 268 | 268  | 253-269 \cr
 \etil_L        |455   | 450 | 449  | 451-452\cr
\etil_R         | 419  | 417 | 418  | 417-419\cr
\nutil_e        |447   | 441 | 441  | 442-445\cr
\nutil_{\tau}   | 395  | 390 | 390  | 387-393 \cr 
\chi_1          | 128  | 120 | 120  | 119-121\cr
\chi_2          |242   | 226 | 225  | 228-228\cr
\chi_3          | 400  | 419 | 415  | 406-420\cr
\chi_4          | 420  | 435 | 431  | 422-436\cr
\chi^{\pm}_1    | 242  | 226 | 225  | 227-228\cr
\chi^{\pm}_2    | 421  | 436 | 432  | 422-436 \cr
h               | 116  | 116 | 116  | 114-116\cr
H               | 370  | 386 | 385  | 355-367\cr
A               | 371  | 388 | 387  | 355-367\cr
H^{\pm}         | 381  | 397 | 396  | 366-379 
\endtable}

\centerline{{\it Table~7:\/} Sparticle masses    
(in $\GeV$) for the SPS4 point}        
\medskip

\subsec {Benchmark point SPS 5}
\medskip
\vbox{
\begintable
{\rm mass} |  1{\rm loop} | 2 {\rm loops} | 3 {\rm loops} | {\rm AKP} \cr
 \gtil |  743 | 729 |727 | 719-729 \cr
  \ttil_1  | 653| 654| 646| 629-651\cr 
  \ttil_2  | 265| 278| 263| 258-280\cr   
  \util_L   |684 | 677 |668 | 676-685 \cr
 \util_R  | 658 | 656|646| 655-660 \cr   
 \btil_1  |563  | 563 |554   | 554-567\cr
  \btil_2  |654  |653 | 643 | 630-656\cr
 \dtil_L  | 688 | 681| 673 |681-689 \cr
  \dtil_R  | 656 | 655|645 | 653-658\cr
  \tautil_1  |264  |259 | 258  | 259-262\cr
 \tautil_2  | 186  |182 |183   | 182-184\cr
 \etil_L  | 263| 257| 257 | 258-261\cr
  \etil_R  | 195 | 192|193 | 192-194\cr
 \nutil_e   | 251|  245| 245 | 246-249\cr
 \nutil_{\tau}   |249 | 243 | 243 | 244-247\cr
 \chi_1    | 128 | 120| 120| 119-120\cr
 \chi_2   |247 |229  | 228 | 230-236\cr
 \chi_3   |608 |626  | 621 | 626-631\cr
 \chi_4   | 621| 637 | 632 | 637-641\cr
 \chi^{\pm}_1   |247 | 229 |228  | 230-236\cr
 \chi^{\pm}_2   |620 | 637 | 632 | 636-641\cr
 h   | 117| 118 |118  | 116-122 \cr
 H   | 667|682  | 676 | 681-694\cr
 A   | 667| 682 | 677 | 682-690\cr
 H^{\pm}   |672 |687  |681  | 687-698
\endtable}
\centerline{{\it Table~8:\/} Sparticle masses    
(in $\GeV$) for the SPS5 point with $m_t = 178\GeV$}        
\medskip

This point differs from the previous ones in having a large 
value of the $A$-parameter.
The  contributions of $\mu, A$   
to the off-diagonal term in the stop mass matrix have the same sign,
and the magnitude of $A$ is large, resulting in a light stop.
For this point we have calculated both using in Table~8 $m_t = 178\GeV$ (as for 
the previous tables) and for comparison in Table~9 with 
$m_t = 174.3\GeV$. This illustrates the sensitivity to the 
input $m_t$, with the light stop changing over $20\GeV$ due to  this small 
change in $m_t$.
\vskip3em
\vbox{
\begintable
{\rm mass} |  1{\rm loop} | 2 {\rm loops} | 3 {\rm loops} | {\rm AKP} \cr
 \gtil  |  743 | 729 |727 | 718-728 \cr   
  \ttil_1  | 652| 653| 645| 628-649\cr
  \ttil_2  | 243| 257| 240| 232-258\cr
  \util_L   |684 | 677 |668 | 676-684 \cr
 \util_R  |658 | 656|646| 653-660 \cr
 \btil_1  |561  | 560 |551   | 551-564\cr
  \btil_2  |654  |653 | 643 | 629-655\cr
 \dtil_L  | 689 | 681| 673 | 680-689\cr
  \dtil_R  | 656 | 655|645 | 651-658\cr
  \tautil_1  |264  |259 | 258  | 258-262\cr
 \tautil_2  | 186  |182 |182   | 182-184\cr
 \etil_L  | 263| 257| 257 | 258-260\cr
  \etil_R  | 195 | 192|192 | 192-194\cr
 \nutil_e        | 251|  245| 245 | 246-249\cr
 \nutil_{\tau}   |249 | 243 | 243 | 244-246\cr
 \chi_1    | 128 | 120| 120| 119-121\cr
 \chi_2   |247 |229  | 228 | 230-236\cr  
 \chi_3   |615 |632  | 628 | 632-637\cr
 \chi_4   | 628| 644 | 639 | 643-646\cr  
 \chi^{\pm}_1   |247 | 229 |228  | 230-236\cr
 \chi^{\pm}_2   |627 | 643 | 639 | 643-646\cr
 h   | 115| 115 |115  | 112-119 \cr
 H   | 674|688  | 683 |687-693  \cr
 A        | 674| 688 | 683 | 689-693\cr
 H^{\pm}   |679 |692  |687  |  694-702
\endtable}
\centerline{{\it Table~9:\/} Sparticle masses    
(in $\GeV$) for the SPS5 point with $m_t = 174.3\GeV$}        
\medskip

\subsec {Benchmark point SPS6}

This is a point with un-unified gaugino masses so 
we are unable to compare with Ref.~\sabine.
We instead use the paper by Ghodbane and Martyn (GM), Ref.~\GKG, 
which also compares the results for 
various programs (Isajet, Susygen and Pythia).
The results for these three programs are reasonably consistent 
with each other; this is due to some extent, however,  to the fact that 
the Isajet gauge unification outputs are used as 
inputs for the other two programs; in our table we show only the Isajet
predictions.  
Agreement with our results  is less impressive; however 
we should notice that  Ref.~\GKG\ uses an earlier version of Isajet (7.58) 
than Ref.~\sabine. Thus if we 
return to SPS1a and  compare the Isajet~7.58 prediction for 
the gluino mass ($595\GeV$) with the Isajet~7.69 one of 
$612\GeV$ obtained from Ref.~\sabine, we can anticipate that for SPS6
the more recent Isajet would give results more consistent with 
our (two-loop) ones, making the reasonable assumption that the newer 
version will give, for example, 
 a higher gluino mass prediction for SPS6 as well. 
Our results for SPS6 are given in Table~10.
\vfil\eject
\vbox{
\begintable
 {\rm mass} | 1{\rm loop}  |2{\rm loops}| 3{\rm loops} | {\rm GM} \cr
 {\tilde g}     | 744  | 726 | 724  | 708\cr
 \ttil_1        | 686  | 681 | 673  | 661\cr
 \ttil_2        | 498  | 496 | 488  | 476\cr
 \util_L        | 684  | 674 | 665  | 639 \cr
\util_R         | 665  | 659 | 650  | 628 \cr
\btil_1         | 620  | 613 | 605  | 589\cr
 \btil_2        | 657  | 652 | 643  | 624\cr
 \dtil_L        | 689  | 679 | 670  | 644 \cr
\dtil_R         | 658  | 653 | 644  | 622 \cr
\tautil_1       | 278  | 271 | 271  | 270\cr
 \tautil_2      | 235  | 229 | 229  | 228 \cr
 \etil_L        | 274  | 266 | 266  | 265\cr
\etil_R         | 243  | 238 | 238  | 237\cr
\nutil_e        | 261  | 253 | 253  | 252  \cr
\nutil_{\tau}   | 260  | 252 | 252  | 252 \cr 
\chi_1          | 201  | 190 | 190  | 189\cr
\chi_2          | 239  | 222 | 221  | 218\cr
\chi_3          | 399  | 419 | 414  | 399\cr
\chi_4          | 425  | 439 | 435  | 420\cr
\chi^{\pm}_1    | 237  | 220 | 219  | 215\cr
\chi^{\pm}_2    | 423  | 438 | 434  | 419 \cr
h               | 115  | 115 | 115  | 115\cr
H               | 469  | 481 | 477  | 464\cr
A               | 469  | 481 | 477  | 463\cr
H^{\pm}         | 477  | 489 | 484  | 470 
\endtable}

\centerline{{\it Table~10:\/} Sparticle masses    
(in $\GeV$) for the SPS6 point}        
\medskip

\subsec{Discussion}

A clear feature of the results is that the corrections 
due to two and three-loop running can be quite large for squarks,
but are typically smaller for weakly-interacting particles. 
In particular the 
light CP-even Higgs mass is very stable. 
The large three-loop 
$\alpha_3$ corrections stem mainly from the $M_3^2$ contributions 
to the three-loop $m^2$ $\beta$-functions; note that for the only MSUGRA 
point such that $m_0 > m_{\half}$, i.e. SPS2, the three-loop correction 
to the squark masses is {\it smaller\/} than the two-loop one. 

Generally speaking we would anticipate  that for  regions of parameter
space where the  three-loop corrections are  comparable to or exceed the
two-loop ones, the four-loop ones  will be at least as large. This
suggests that  we are already at three loops approaching the asymptotic 
region for the $\beta$-functions.  So it  appears that squark mass
predictions  with an accuracy greater than a few per cent will not be
possible using perturbation theory.

Overall our results agree  reasonably well with those of existing
programs~\sabine.  One place where we have a significant difference is
for the  $H,A,H^{\pm}$ results for SPS4. This is a large $\tan\beta$
point; however  our results for the $b$-squark and $d$-squark masses
(which  one would expect to be sensitive to large $\tan\beta$) agree
quite well,  so for the moment we have no explanation for this
discrepancy.

\newsec{The Semi-perturbative Region}

The addition of additional matter representations
in complete $SU_5$ multiplets does not affect gauge 
unification (and the unification scale) at one loop. 
Beyond one loop this is no longer the case, and 
increasing the amount of matter relevant to the running analysis 
requires the presumption of larger threshold corrections 
at the unification scale in order to restore gauge unification; 
one is thus  forced to argue that the success 
of gauge unification in the MSSM is coincidental
\foot{Historically gauge unification was implemented by 
using $\alpha_3 (M_Z)$ as an input and computing $\sin^2\theta_W$, although
the latter was more accurately measured, because 
$\sin^2\theta_W$ varies very slowly with $\alpha_3 (M_Z)$, and 
conversely (of course) $\alpha_3 (M_Z)$ varies rapidly as a function 
of $\sin^2\theta_W$.
The current experimental results for $\alpha_3 (M_Z)$ 
already require us to 
suppose the existence of {\it some\/} high scale radiative corrections 
in the MSSM;
but the fact remains that things get worse as we add more matter\glr.}.  
\smallskip
\epsfysize= 4in
\centerline{\epsfbox{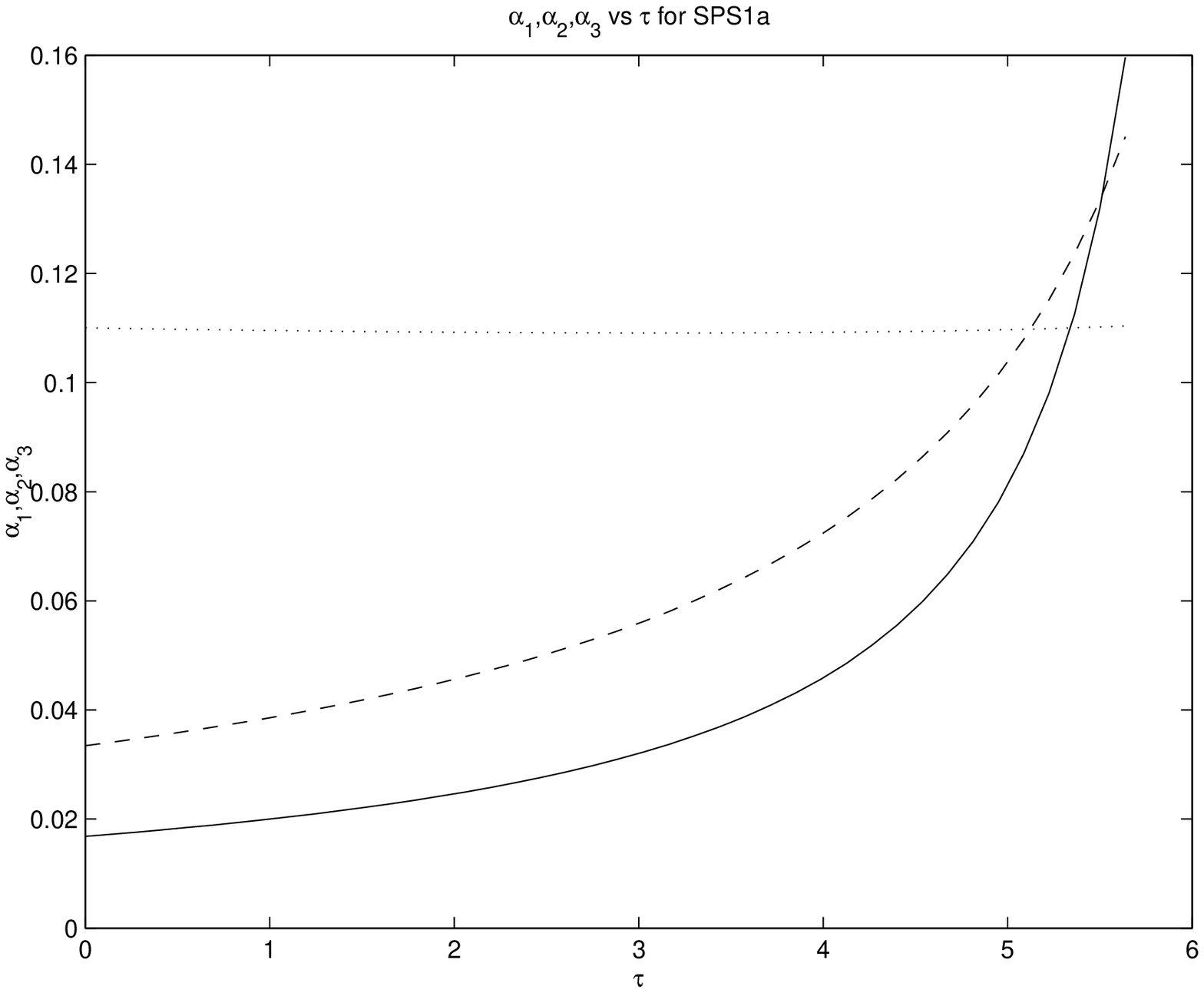}}
\inparg
{\it \noindent Fig.1:
Gauge coupling unification for $n_{10} = 1.7$.
Solid, dashed, and dotted lines correspond to 
$\alpha_1,\alpha_2,\alpha_3$ respectively.}
\medskip
\outparg

In Fig.~1 we show the evolution of the gauge couplings 
$\alpha_i = g_i^2/(4\pi)$ 
for $n_{10} = 1.7$, using three-loop $\beta$-functions 
for all couplings. (As remarked in Ref.~\kmr, 
the mass scale of these additional multiplets being unknown it makes
sense to  parametrise their effects by taking $n_5, n_{10}$ to be
continuous variables.) 
The couplings are plotted 
against $\tau = \frak{1}{2\pi}\ln(Q/M_Z)$; evidently we are still in the 
perturbative regime. The input parameters at $M_Z$ correspond 
to a typical supersymmetric mass spectrum;
specifically, the Benchmark point SPS1a. One sees clearly the 
need for large corrections to restore gauge  unification. 

We gave a number of examples of the effect of additional matter 
on the sparticle spectrum predictions in a previous paper\jjk; here 
we contrast the effect on the first and third generation squark masses. 
Thus in Fig~2 we plot, for the SPS5 point, 
the ratio of the $\util_L$ and gluino 
masses against $n_{10}$ for 
$n_5 = 0$; as already noted in Ref.~\kmr, the mass increases with $n_{10}$. 
It is interesting that the effect of the three-loop correction 
to this ratio almost precisely cancels the two-loop correction,
for all $n_{10}$.
We contrast this with Fig~3 where we show the behaviour of the light 
stop mass for the same SPS point; in this case the 
ratio decreases smoothly, and the three-loop correction only 
cancels the two-loop one at $n_{10} = 0$.
For the SPS5 point the electroweak vacuum fails around $n_{10} = 0.48$.
(The change in this value and in Fig.~3 from our previous paper\jjk\ 
is due to the change in the input top pole mass, and to an improved 
treatment of the Higgs potential minimisation.)

In Fig.~4 we plot the light CP-even Higgs mass  
for SPS1a as a function of $n_{10}$ (for $n_5 =
 0$). 
We see that it is fairly stable both with respect to loop corrections and 
the addition of extra matter. In the case of SPS1a the electroweak 
vacuum fails at around $n_{10} = 1.8.$
\smallskip
\epsfysize= 3in
\centerline{\epsfbox{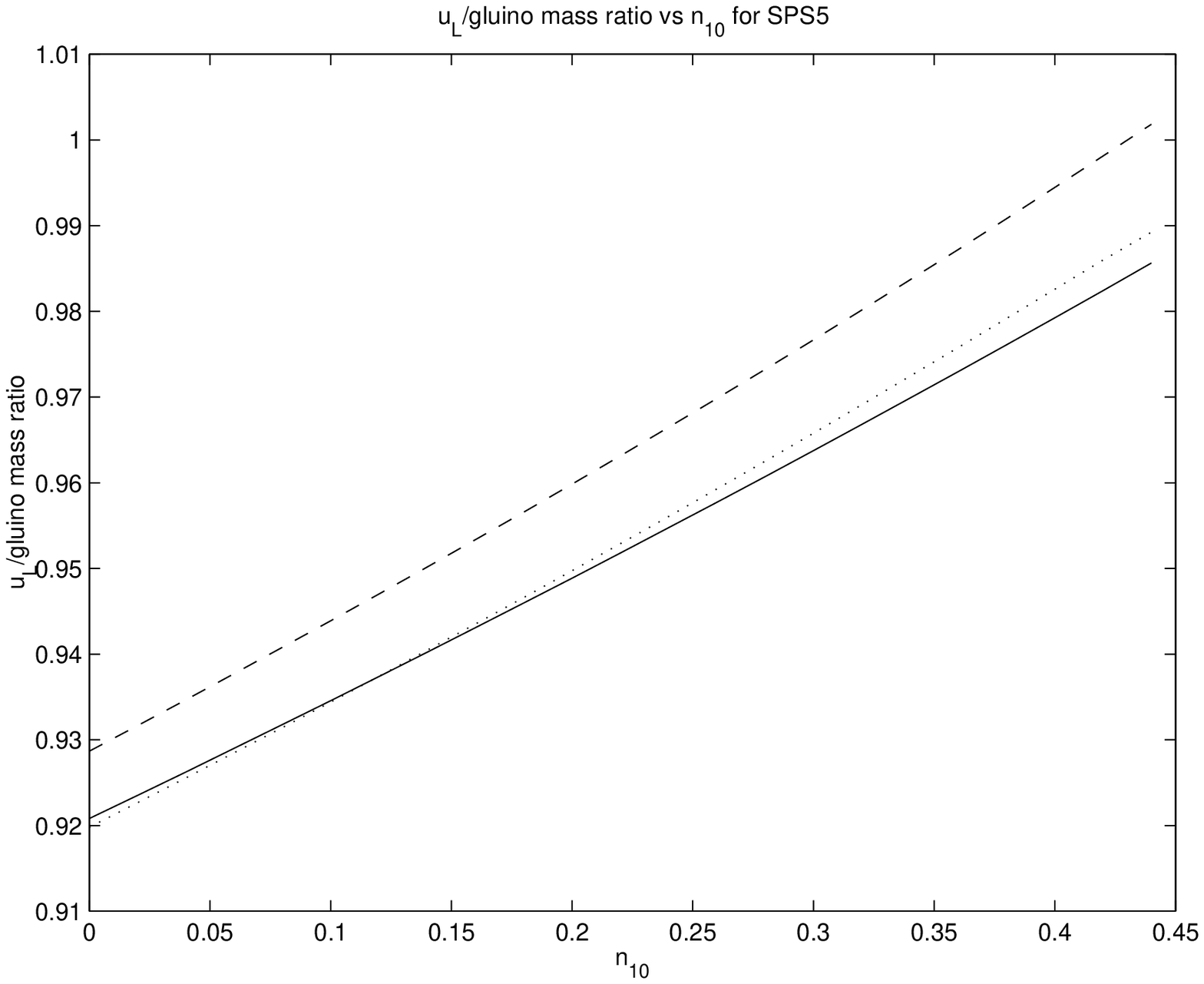}}
\inparg
{\it \noindent Fig.2:
Plot of the $\util_L/gluino$ mass ratio against $n_{10}$
for SPS5. Solid, dashed and dotted lines correspond to one, two and
three-loop running respectively.}
\medskip
\outparg
\smallskip
\epsfysize= 3in
\centerline{\epsfbox{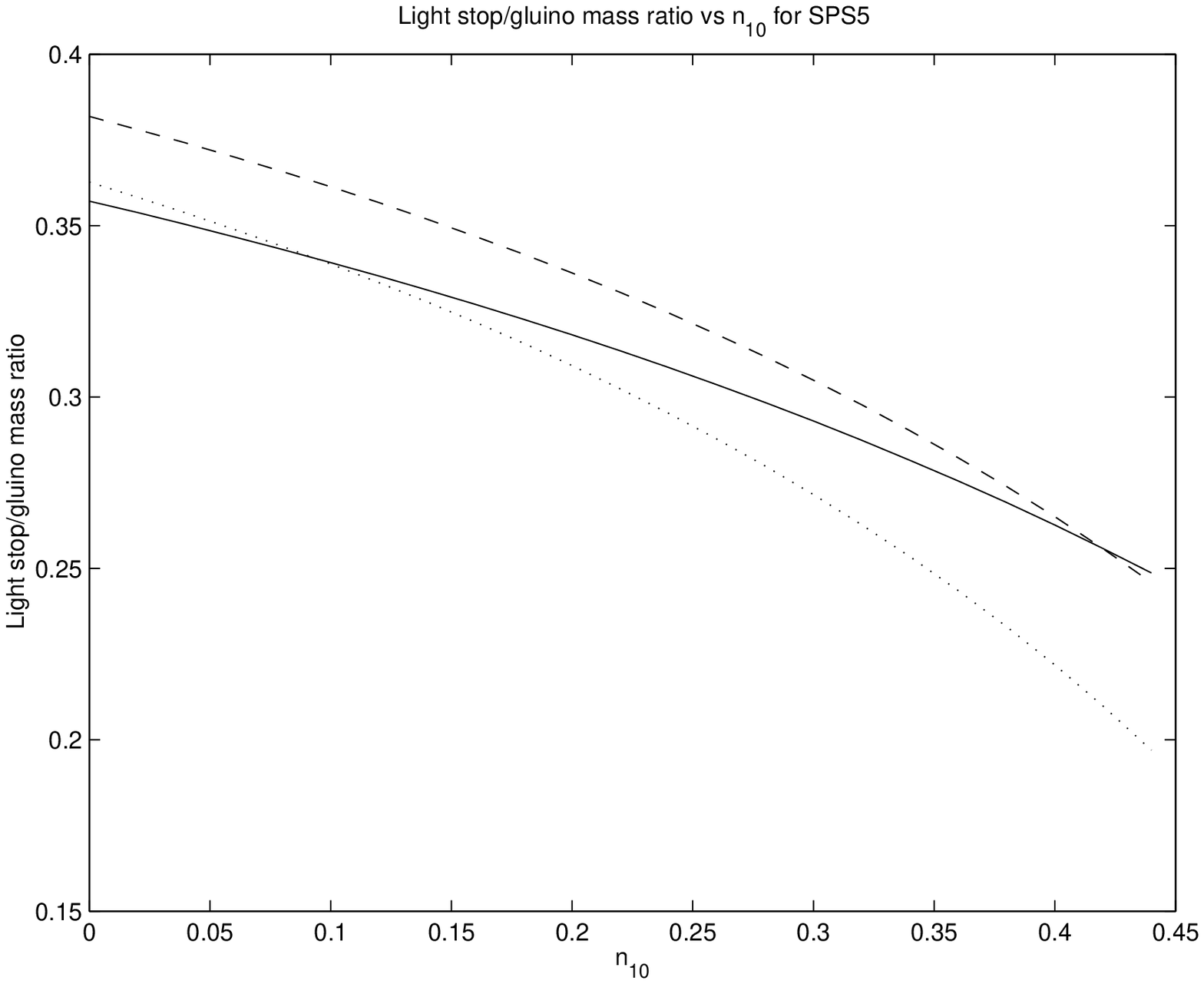}}
\inparg
{\it \noindent Fig.3:
Plot of the light stop/gluino mass ratio against $n_{10}$
for SPS5. Solid, dashed and dotted lines correspond to one, two and
three-loop running respectively.}
\medskip
\outparg

\smallskip
\epsfysize= 3.0in
\centerline{\epsfbox{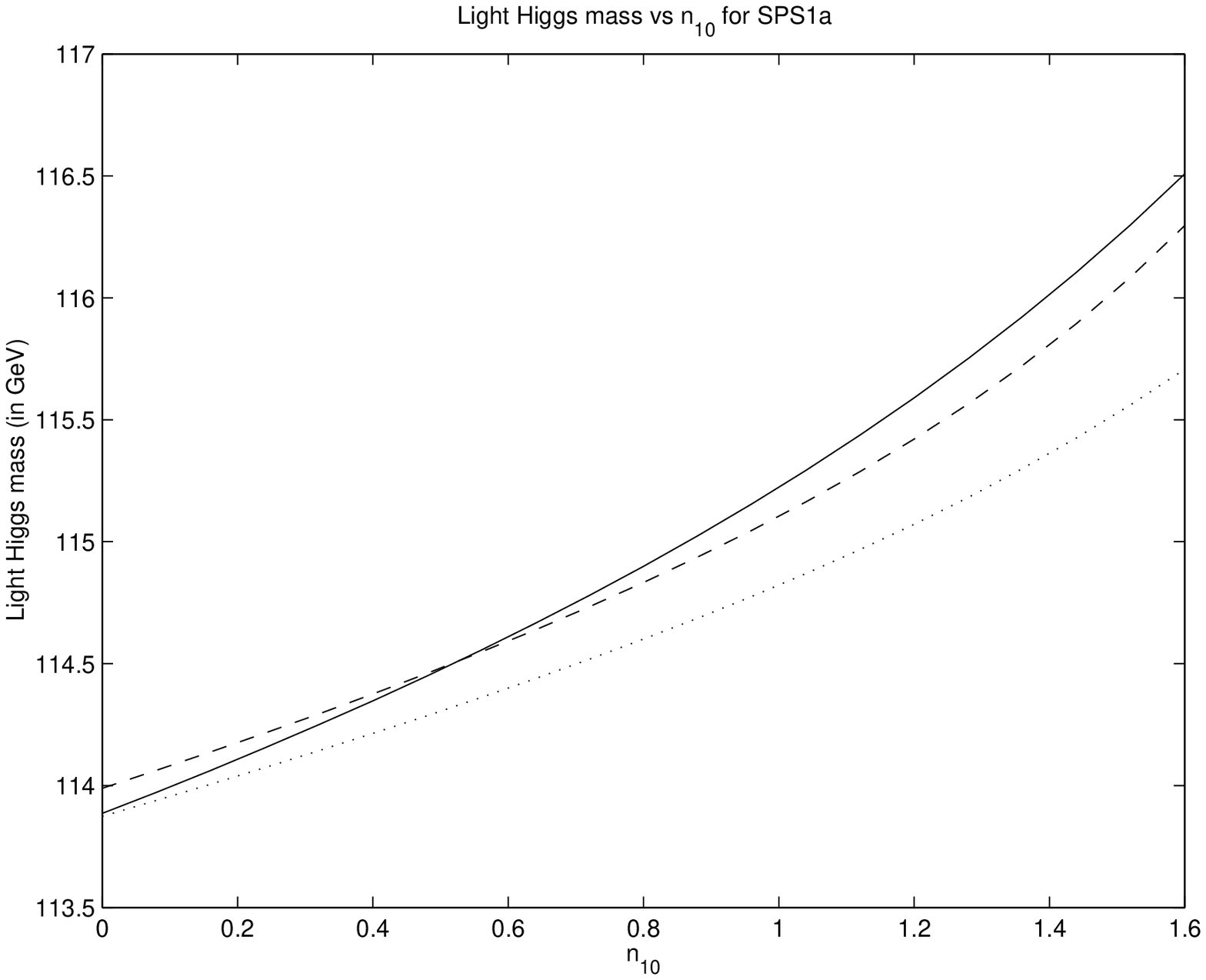}}
\inparg
{\it \noindent Fig.4:
Plot of the light CP-even Higgs mass against $n_{10}$ for SPS1a.
Solid, dashed and dotted lines correspond to one, two and
three-loop running respectively.}
\smallskip
\outparg

\newsec{Conclusions}

We have extended typical detailed running coupling analyses for  the
MSUGRA MSSM SPS benchmark points to incorporate three-loop
$\beta$-function corrections for the running masses and couplings. We
compare our results to those obtained by existing  programs using 
two-loop running.  The spread in the results from these programs  is
probably due to a mixture of program errors and genuine  theoretical
uncertainties such as the choice of scale appropriate for the 
evaluation of the pole mass. Presumably  over time the results used by
these programs will converge;  we would argue that a more reliable 
estimate of the ultimate theoretical error in these  spectrum
calculations is currently provided by the difference between our two and
three-loop calculations,  as opposed to the spread in the various
available two-loop results.

Generally speaking the effect of the three-loop running corrections 
is small for weakly-interacting particles but larger for the squark 
masses. For the light stop mass at the SPS5 point, 
we see an $8\%$ effect, but more typically the effect is 
between $1\%$ and $2\%$. This appears to us to represent a fundamental limit 
on the theoretical precision of squark mass theoretical 
predictions. 

Finally we show how additional matter in $SU_5$ multiplets can affect 
the sparticle spectrum; more dramatically as the ``semi-perturbative 
unification'' regime~\kmr\  is approached.

\medskip

\bigskip\centerline{{\bf Acknowledgements}}\nobreak

DRTJ was  supported by a PPARC Senior Fellowship,  and  a CERN Research
Associate-ship, and was visiting  CERN while most of this work was done.
AK was supported  by an Iranian Government Studentship.   We thank Ben
Allanach,  Jon Bagger, Ruth Browne, John Gracey, Tilman Plehn and Graham
Ross  for conversations.   
 
\listrefs
\bye